\documentclass[11pt, onecolumn]{article}
\usepackage{geometry}
\usepackage{cite}
\usepackage{hyperref}
\usepackage{amsmath,amssymb,amsfonts}
\usepackage{subfig}
\usepackage{graphicx}
\usepackage{epstopdf}
\usepackage{fancyhdr}
\usepackage{upgreek}
\usepackage{textcomp}
\usepackage{siunitx}
\usepackage{authblk}
\usepackage[version=4]{mhchem}
\usepackage{xcolor}
\date{}

\author[1]{Patrik Gubeljak$^\dagger$}
\author[2,3]{Tianhui Xu$^\dagger$}
\author[3,4]{Lorenzo Pedrazzetti}
\author[3]{Oliver J. Burton}
\author[4]{Luca Magagnin}
\author[3]{Stephan Hofmann}
\author[3]{George G. Malliaras}
\author[1,2,5]{Antonio Lombardo*}
\affil[1]{Cambridge Graphene Centre, Department of Engineering, University of Cambridge, United Kingdom}
\affil[2]{Department of Electronic and Electrical Engineering, University College London, London, United Kingdom}
\affil[3]{Department of Engineering, University of Cambridge, United Kingdom}
\affil[4]{Dipartimento di Chimica, Materiali e Ingegneria Chimica “Giulio Natta”, Politecnico di Milano, Italy}
\affil[5]{London Centre for Nanotechnology, University College London,  United Kingdom}

\begin{document}
\title{Electrochemically-gated Graphene Broadband Microwave Waveguides for Ultrasensitive Biosensing}

\maketitle
\vspace{-1cm}
$^\dagger$ These authors contributed equally to this work. *Corresponding author: \href{mailto:a.lombardo@ucl.ac.uk}{a.lombardo@ucl.ac.uk}

\begin{abstract}

Identification of non-amplified DNA sequences and single-base mutations is essential for molecular biology and genetic diagnostics. This paper reports a novel sensor consisting of electrochemically-gated graphene coplanar waveguides coupled with a microfluidic channel. Upon exposure to analytes, propagation of electromagnetic waves in the waveguides is modified as a result of interactions with the fringing field and modulation of graphene dynamic conductivity resulting from electrostatic gating. Probe DNA sequences are immobilised on the graphene surface, and the sensor is exposed to DNA sequences which either perfectly match the probe, contain a single-base mismatch or are unrelated. By monitoring the scattering parameters at frequencies between 50 MHz and 50 GHz, unambiguous and reproducible discrimination of the different strands is achieved at concentrations as low as one attomole per litre (1 a\textsc{m}). By controlling and synchronising frequency sweeps, electrochemical gating, and liquid flow in the microfluidic channel, the sensor generates multidimensional datasets. Advanced data analysis techniques are utilised to take full advantage of the richness of the dataset. A classification accuracy \textgreater{} 97\% between all three sequences is achieved using different Machine Learning models, even in the presence of simulated noise and low signal-to-noise ratios. The sensor exceeds state-of-the-art sensitivity of field-effect transistors and microwave sensors for the identification of single-base mismatches. 

\end{abstract}

\section{Introduction}

Biosensors capable of identifying non-amplified DNA sequences with high sensitivity and selectivity are essential for applications ranging from fundamental molecular biology to genetic disease diagnosis and precision medicine. Electronic detectors, such as field effect transistors (FETs), are of particular interest as they can combine label-free detection, high sensitivity, small footprint and integrability with conventional electronics for signal processing~\cite{SadiTRAC133}. Graphene attracted significant research and commercial interest for biosensing due to its electrical and chemical properties, high surface-to-volume ratio, biocompatibility, and ease of functionalisation~\cite{BitoAM25,PratTB39}. Exposure to chemical species, such as gases~\cite{Schedin2007}, ionic solutions ~\cite{LiSAB253}, enzymes~\cite{enzyme}, glucose~\cite{ZhanSR5}, large biomolecules~\cite{Bharti2020}, viruses~\cite{covid_graphene}, and bacteria~\cite{bacterium_sensor}, modifies graphene electronic properties, typically as a result of the modulation in the density and scattering rates of charge carriers. By incorporating graphene in a transistor structure, usually referred to as Graphene Field Effect Transistor (GFET)~\cite{novoselov2004electric}, the results of these interactions can be measured on a macroscopic scale, typically by monitoring changes of the charge neutrality point (CNP) in the transfer characteristics~\cite{Schedin2007, LiSAB253, Xu2017_realtime_DNA, Campos2019_attomolar_DNA}. Biomaterials are usually dispersed in a suitable medium, typically an electrolyte buffer solution. When in contact with ionic media, electrical double layers (EDLs), also known as Debye layers, form at the graphene-electrolyte interface, resulting in a large interface capacitance $C_{EDL}$, due to the small thickness (Debye length) of the EDL~\cite{Das2008}. This effect is used in electrochemically-gated GFETs, where the graphene channel is exposed to the ionic solution, and a voltage is applied to a counter electrode, modulating the EDL and, in turn, the charge carrier density in the graphene channel. An EDL also forms at the electrode-solution interface, leading to a capacitance in series with $C_{EDL}$. However, counter-electrodes are usually designed to have areas significantly larger than the graphene channel, resulting in a very large capacitance whose effect is negligible in the series. Under such conditions, the voltage applied to the counter electrode drops almost entirely at the graphene-solution interface, i.e. across the EDL~\cite{ZhanSR5, Xu2017_realtime_DNA, Campos2019_attomolar_DNA}. The total gate capacitance of electrochemically-gated GFETs, therefore, consists of graphene quantum capacitance (due to its finite density of states~\cite{FernPRB75}) $C_Q$ and EDL capacitance in series, i.e. $C_G = [C_Q^{-1} + C_{EDL}^{-1}]^{-1}$~\cite{Xu2017_realtime_DNA}. As $C_G$ is very large, even small changes in the solution are reflected in significant changes in the transistor transfer characteristic, resulting in very high sensitivity and low limits of detection~\cite{Hwang2020_600zM}. To enhance the selectivity of GFET sensors, the graphene surface can be non-covalently functionalised with different groups, which increases the specificity while preserving the electrical conductivity~\cite{ZhanSR5, Xu2017_realtime_DNA, Campos2019_attomolar_DNA}. 

The combination of electrochemically-gated GFETs and surface functionalisation has been applied to identify DNA sequences and single-base mismatches~\cite{Hwang2016_first_DNA, Xu2017_realtime_DNA, Campos2019_attomolar_DNA}. By using a binder molecule (1-pyrenebutanoic acid succinimidyl ester, PBASE) which  attaches non-covalently to the graphene channel, single strands of DNA were immobilised on the GFET. When target DNA strands are introduced to the functionalised sensing surface, the hybridisation with the immobilised probe DNA modifies the potential across the EDL, resulting in shifts of the CNP~\cite{Hwang2016_first_DNA, Xu2017_realtime_DNA, Campos2019_attomolar_DNA}. Xu et al.~\cite{Xu2017_realtime_DNA} used this to distinguish single-base mismatches quantitatively in real-time with a target DNA concentration of 5 nM based on an electrolytically gated GFET array. Campos et al.~\cite{Campos2019_attomolar_DNA} improved their work and demonstrated a limit of detection (LOD) of 25 a\textsc{m} of the lowest target DNA concentration for which the sensor can discriminate between perfect-match sequences and nucleotides having a single base mismatch. 

A major limitation of the sensitivity of FET for biosensing in physiological solutions is the ionic (Debye) shielding, which limits the detection of molecules to only those within the Debye length, i.e. usually between 0.7 and 8 nm, depending on the ionic strength of the solution. This, in turn, reduces the sensitivity, and often complex approaches are required to mitigate the screening~\cite{NakaS362}. However, the Debye shielding only affects devices operated at DC and low frequency and becomes negligible at microwave frequencies as the ionic conductivity vanishes~\cite{Artis2015}. Microwaves interact with matter causing frequency-dependent reorientation of molecular dipoles and translation of electric charges ~\cite{GuarIEEEMM16,Artis2015}. 
Different molecules and compounds are characterised by different relaxation processes (collectively captured by their dielectric permittivity) and interact differently with oscillating electromagnetic fields~\cite{GuarIEEEMM16}. Microwave sensors use such interaction to identify or discriminate different analytes, and have been successfully used to identify cancer cells~\cite{GrenIEEEMTT61}, volatile compounds in breath~\cite{SchmIEEEMTT65}, study antibiotic resistance in bacteria~\cite{JainSR11} and electroporation in human epithelial cells~\cite{TamrIEEEMB3}. Different types of sensors have been reported, including reflectometers, resonators, interferometers, and waveguides~\cite{GuarIEEEMM16}. Waveguide sensors, such as coplanar waveguides (CPWs), are of particular interest as they combine broadband operation with simple design, ease of miniaturisation and integrability with conventional planar technology and microfluidics~\cite{GuarIEEEMM16, Artis2015}. In a CPW, part of the field extends outside of the circuit due to incomplete shielding of the conductors~\cite{GuarIEEEMM16,Artis2015} and therefore interacts with analytes deposited on the waveguide surface. Yang et al.~\cite{YANG2012349} developed a multilayered polymeric radio frequency (RF) sensor for DNA sensing using a CPW sensing surface, which reached a LOD of target DNA of 10 pM through DNA hybridisation, and Kim et al.~\cite{KIM2013362} proposed an RF biosensor based on an oscillator at 2.4 GHz and obtained an estimated LOD of about 1 ng/mL (114 p\textsc{m}). 

Graphene is of particular interest for RF and microwave sensing owing to its good conductivity and field effect tunability~\cite{schwierz2010graphene,gubeljak_Memea2022,zhang2022dielectric}. Moreover, its AC conductivity is frequency-independent and equal to DC conductivity for frequencies up to $\approx 500$ GHz~\cite{Awan2016_RF_conductivity}. This unique combination of properties has been used to demonstrate proof-of-concept electrolytically-gated waveguide sensors, capable of identifying completely complementary DNA strands and generating multidimensional datasets by independently controlling gate voltage and frequency~\cite{gubeljak_Memea2022}. Recently, Zhang et al.~\cite{zhang2022dielectric} reported a GFET operated around its resonant frequency (i.e., 1.83 GHz) in reflectometry mode, achieving a LOD of 1 n\textsc{m} for the detection of streptavidin, an extensively used protein. 

Machine learning (ML) techniques play key roles in the field of biological sequencing, including DNA, RNA, and protein~\cite{10.1093/bib/bbx165}. However, there is not much previous work using ML to analyse raw microwave signals after being exposed to biological samples~\cite{D1QM00665G}. ML regression models and Neural Networks were used on the reflection and transmission coefficients from electrically-small dipole sensors~\cite{bamatraf2022noninvasive} and open-ended coaxial probes~\cite{turgul_permittivity_2018} and achieved either a direct prediction of aqueous glucose solution concentrations or a prediction of the permittivity of glucose solutions. Nevertheless, the authors are not aware of work that applies ML to broadband miniaturized on-chip microwave sensors for biomaterial sensing at the
time of writing. Regarding single-base-mismatch DNA detection, Principal Component Analysis (PCA) and Quadratic Discriminant Analysis (QDA) were applied to Terahertz spectral data and achieved a classification rate of 90.3\% in the prediction set of four single-base-mismatch DNA oligonucleotides at a concentration of $38.85$ $\mu$\textsc{m}~\cite{Tang:20}. 

Here we present a novel DNA sensor consisting of electrochemically-gated graphene CPWs coupled with a microfluidic channel. The sensor harnesses the combined effect of dynamic conductivity modulation in graphene resulting from (chemical) electrostatic doping and modification of wave propagation resulting from the interaction of the fringing field with the analyte. The two effects occur simultaneously, leading to a unique double sensing mechanism that combines two traditionally separate sensing approaches, i.e. field effect transistor sensing and microwave dielectric spectroscopy. By immobilising probe DNA sequences on the graphene surface, the waveguide scattering parameters are studied when the sensor is exposed to DNA sequences either perfectly matching the probe (pmDNA) or containing a single-base mismatch (smDNA) or unrelated (uDNA). Unambiguous and reproducible discrimination of single-base mismatch target strands is achieved at DNA concentrations as low as 1 attomole per litre (1 a\textsc{m}). Multidimensional datasets are obtained by controlling and synchronising frequency sweeps, electrochemical gating, and liquid flow in the microfluidic channel. Such rich datasets are analysed using different ML methods, achieving a classification accuracy \textgreater{} 97\% between pmDNA, smDNA, and uDNA, even in the presence of simulated Gaussian noise. 

\begin{figure}[h]
 
\centerline{\includegraphics[width=15cm]{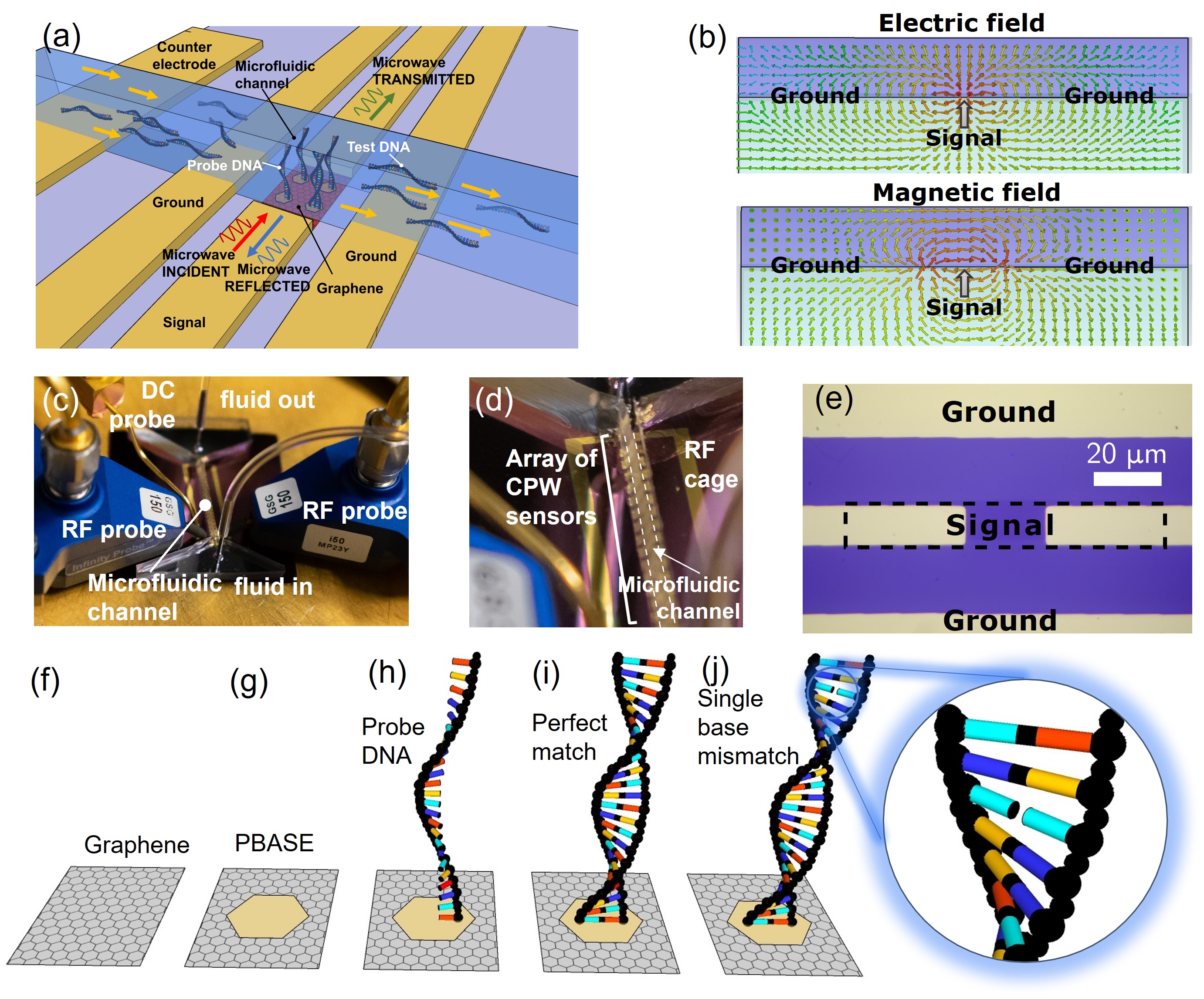}}
\caption{(a) Illustration of the design and components of the device. (b) Cross-section of the simulated electric field and magnetic field in the quasi-TEM mode perpendicular to the wave propagation vector. The light green part is the Si substrate. The light purple part is the material box of analytes that interact with the fringing field. (c) Measurement setup ready for measurement. (d) Close-up of the devices enclosed in the microfluidic channel and the surrounding gate electrode, doubling as an RF cage. (e) Optical micrograph of the graphene section in the central signal conductor of the waveguide, with the graphene outlined in the black dashed line. The overlap between the dashed line and the metal is the area of the graphene-metal contacts. (f) - (j) Conceptual illustration of different chemical functionalisation and measurement stages for DNA detection.}
\label{cpw_sketch}

\end{figure} 

\section{Results and Discussion}

\subsection{Sensor design and sensing principle}

The DNA sensor consists of a graphene channel integrated within a CPW and coupled with a microfluidic channel. The structure of the device is schematically shown in \textbf{Figure~\ref{cpw_sketch} (a)}. Graphene is deposited onto high-resistivity Si substrates covered in 300 nm of \ce{SiO2} and integrated into the signal track of a metallic CPW, whereas the ground conductors are entirely metallic. Arrays of sensors having different graphene lengths (ranging from 10 to 25 $\mu m$) are fabricated on the same chip and share the same microfluidic channel. A planar gold counter-electrode having dimensions significantly ($>700$ times) larger than the graphene channels is fabricated on the chip, enabling electrolytic gating similar to DC sensors~\cite{Xu2017_realtime_DNA, Campos2019_attomolar_DNA}. This structure is designed to achieve a double sensing mechanism. First, electromagnetic waves propagating in the waveguide interact with the liquid in the microfluidic channel via the fringing field, i.e. the portion of the electric and magnetic field extending outside of the waveguide due to incomplete shielding of the conductors~\cite{GuarIEEEMM16}, as shown in \textbf{Figure~\ref{cpw_sketch} (b)}. The choice of high frequencies (50 MHz - 50 GHz) ensures robustness against Debye screening, which degrades sensitivity in DC and low-frequency sensors exposed to ionic solutions such as PBS~\cite{GuarIEEEMM16, zhang2022dielectric}. Second, similar to GFET sensors, the graphene's (DC and AC) conductivity is modified by the proximity with the liquid via the electrostatic effect resulting from the formation of an EDL at the graphene/solution interface~\cite{Xu2017_realtime_DNA, Campos2019_attomolar_DNA}. This further influences wave propagation, enhancing the response of the sensor. This dual sensing mechanism is fully captured by the waveguide scattering ($S$) parameters, which represent the ratios of the transmitted ($S_{21}$ parameter) or reflected ($S_{11}$ parameter) voltage wave and a known "stimulus" wave launched in the waveguide. \textbf{Figure~\ref{cpw_sketch} (c)} shows the chip mounted on a probe station setup, while \textbf{Figure~\ref{cpw_sketch} (d)} depicts a closer view of the system, showing individual devices and the microfluidic channel. \textbf{Figure~\ref{cpw_sketch} (e)} shows an optical micrograph of the fabricated device prior to the deposition of the microfluidic channel. The dashed area corresponds to the graphene layer, including the part underneath the contact areas. DNA sequences are immobilised onto the graphene surface by using PBASE as a linker molecule, following the protocol from Ref.~\cite{Campos2019_attomolar_DNA}. The pyrene group of PBASE binds non-covalently to graphene via $\pi-\pi$ orbital stacking, whereas its succinimide group binds to the 5' end of a purposely modified single-stranded DNA, which is used as the probe, i.e. as the complementary sequence of the DNA to be detected. In order to saturate any non-reacted succinimide group, the sensor is exposed to an ethanolamine solution. The sensor containing the probe DNA is then exposed to dispersions containing either pmDNA or smDNA or uDNA, both dispersed in 1\% PBS at a concentration of one a\textsc{m} per litre (1 a \textsc{M}). The 1\% PBS concentration is chosen for consistency with previous studies on GFET-based DNA sensors~\cite{Xu2017_realtime_DNA,Campos2019_attomolar_DNA}, and corresponds to a Debye length of $\approx 7$ nm~\cite{NakaS362}, matching the length of the hybridised DNA. 

\textbf{Figure~\ref{cpw_sketch} (f) - (h)} summarises the functionalisation of the sensing surface to immobilise the probe single DNA strand on the graphene surface for target DNA hybridisation, whereas (i) and (j) show hybridisation of the probe DNA with a pmDNA or smDNA, respectively. 
The isoelectric point of DNA is at pH $\approx 5$~\cite{Sherbet1983}, while the pH of our dispersion is 7.2, resulting in the oligonucleotides being negatively charged~\cite{Campos2019_attomolar_DNA}. Upon hybridisation, i.e. when pmDNA or smDNA binds with the probe DNA by forming bonds between complementary bases (cytosine-guanine and adenine-thymine), the additional negative charge modifies the electrical double layer formed at the graphene-solution interface, leading to a lowering of the graphene's Fermi level (equivalent to p-doping) and an increase of the scattering time $\tau$~\cite{lin2013label}. This results in a modulation of graphene dynamic conductivity, which can be described by the Kubo formula for intraband transitions~\cite{Awan2016_RF_conductivity}:

\begin{equation}
\begin{gathered}
\sigma_{\mathrm{intra}}\left(\omega,E_{F},\tau,T\right)=\frac{\mathrm i{e}^2k_BT}{\pi\hbar^2\left(\omega+i\tau^{-1}\right)}\Biggl[\frac{E_{F}}{k_BT} + 2\mathrm{ln}\biggl(1+\mathrm{e}^{-\frac{E_{F}}{k_BT}}\biggr)\Biggr]
\end{gathered}
\label{equ_intra}
\end{equation}
\noindent where: $\omega=2\pi f$ is the angular frequency, $E_F$ is the Fermi energy, $\tau$ is the scattering time (assumed to be independent of energy), $T$ is the temperature expressed in Kelvin, $e=1.6\cdot10^{-19}\hspace{0.1cm}$\, C is the electron charge, $\hbar=\frac{h}{2\pi}$ is the reduced Planck’s constant, and 
$k_B=1.38\cdot10^{-23}\hspace{0.1cm}\mathrm{\frac{J}{K}}$ is the Boltzmann’s constant.\\

The double helix DNA resulting from hybridisation with smDNA has different electrical and mechanical properties compared to the one with the pmDNA target. In particular, single-base mutations disrupt the long-range electron transfer within the DNA double-helix~\cite{WongAC78}, which results in different responses to the fringing field of the propagating wave and different modifications of the graphene conductivity. Differences in capacitance corresponding to such mutations are also reflected in the electrostatic gating via the EDL~\cite{Xu2017_realtime_DNA, Campos2019_attomolar_DNA}. On the other hand, uDNA is not expected to bind with probe DNA and form double helix DNA. Wave propagation in the graphene waveguide captures these differences, leading to a novel sensing paradigm combining field effect sensing and dielectric spectroscopy. 

\subsection{DC measurements}
\begin{figure*}[!h]
\centering
\subfloat[]{\includegraphics[width=0.3\linewidth]{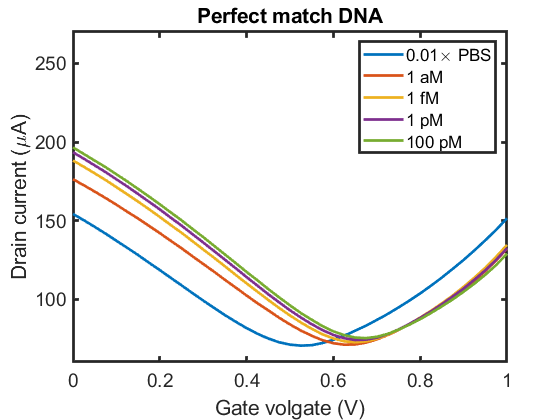}
\label{fig:transfer_curve_pm}}
\hfil
\subfloat[]{\includegraphics[width=0.3\linewidth]{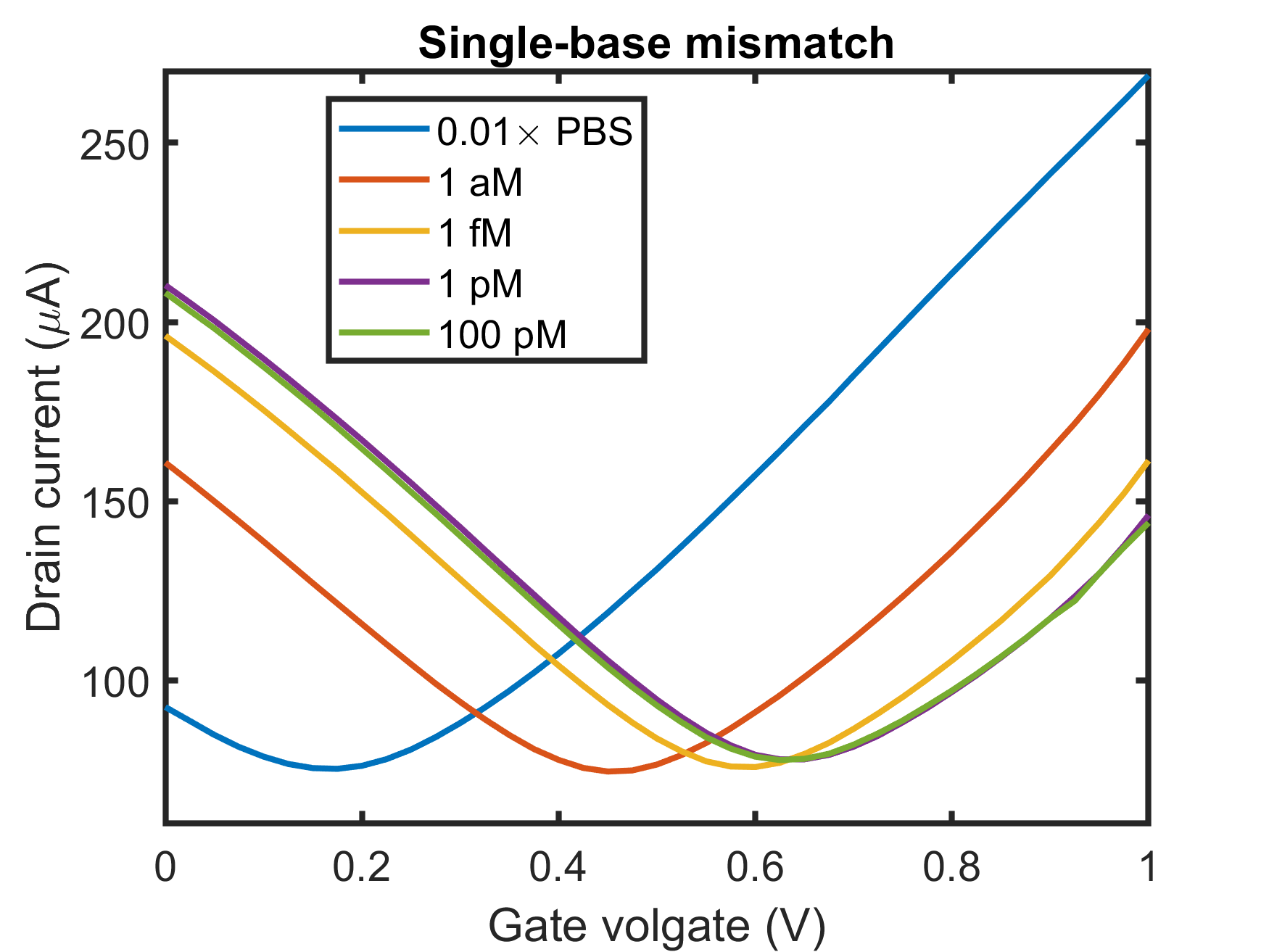}
\label{fig:transfer_curve_sm}}
\hfil
\subfloat[]{\includegraphics[width=0.3\linewidth]{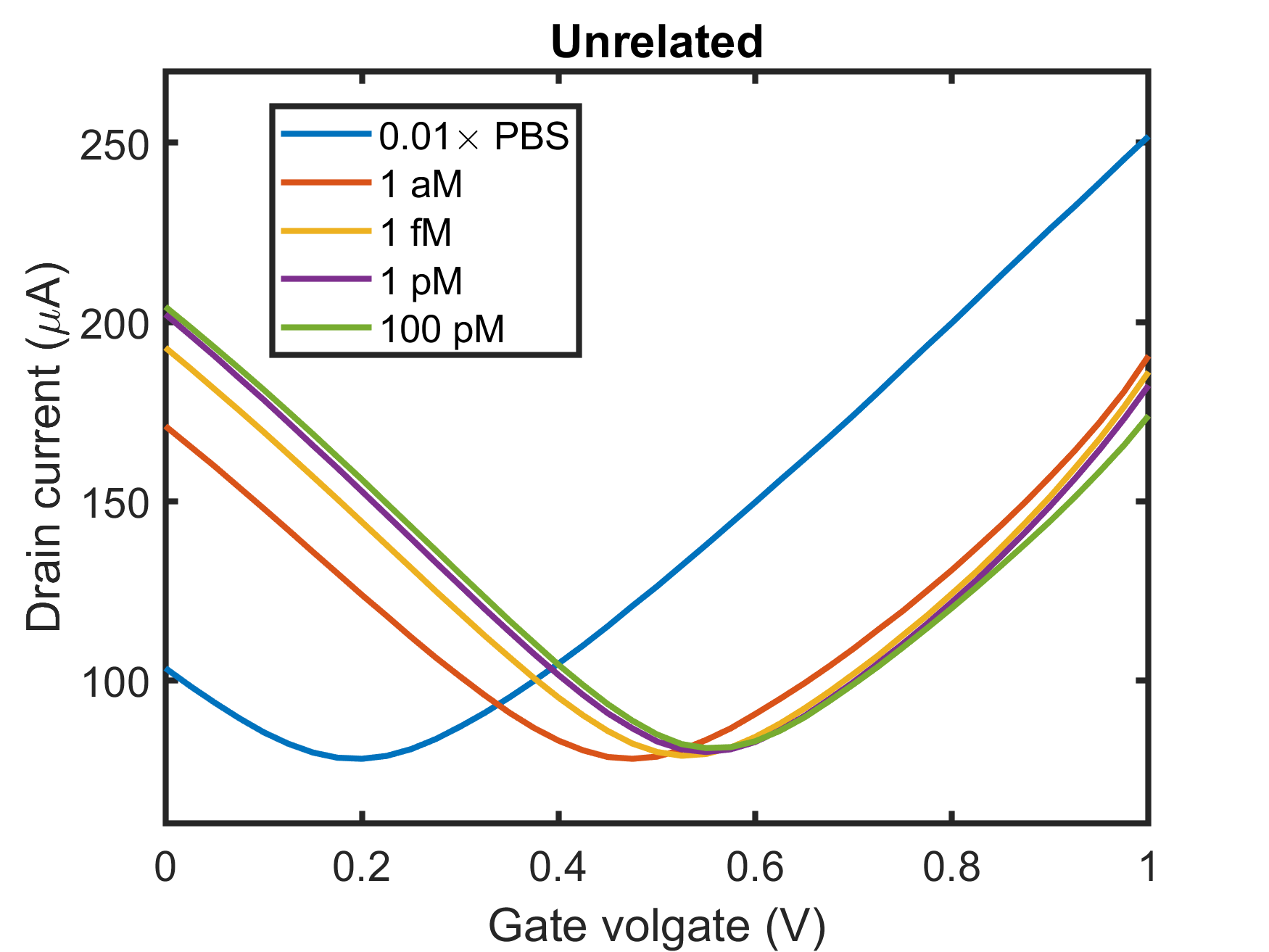}
\label{fig:transfer_curve_u}}
\caption{DC transfer characteristic of the graphene sensors upon hybridization of the probe DNA with (a) pmDNA, (b) smDNA, and (c) uDNA at different DNA concentrations. $0.01 \times$ PBS corresponds to the sensor exposed to buffer solution without DNA.}
\label{fig:transfer_curves}
\end{figure*}

In order to benchmark the sensors against GFET DNA detectors previously reported~\cite{Xu2017_realtime_DNA, Campos2019_attomolar_DNA}, the devices are first tested at DC by using the planar counter electrode as an electrolytic gate at DNA concentrations ranging from 1 a\textsc{m} to 100 p\textsc{m}. \textbf{Figure~\ref{fig:transfer_curves}} shows the DC transfer characteristic of a representative sensor exposed to buffer only and after exposure to a solution of pmDNA (\textbf{Figure~\ref{fig:transfer_curves}}a), smDNA (\textbf{Figure~\ref{fig:transfer_curves}}b) and uDNA(\textbf{Figure~\ref{fig:transfer_curves}}c) at concentrations varying 1 a\textsc{m} to p\textsc{m}. The devices show the typical right-shift of the charge neutrality point (CNP) with increasing concentration for both pmDNA and smDNA~\cite{Xu2017_realtime_DNA, Campos2019_attomolar_DNA}, with a possible onset of saturation at $\approx 10^{-12}$ \textsc{m} and limit of detection $<1$ a\textsc{m}. The right-shift of the CNP for both pmDNA and smDNA confirms that the modulation of graphene conductivity is caused by electrostatic gating via accumulation of charges at the EDL rather than charge transfer between DNA and graphene, which would instead result in a left-shift~\cite{ChenBB156}. The DC characteristics of the sensors evidence that the sensor responds to pmDNA and smDNA at all the concentrations considered, matching state-of-the-art planar GFET sensors~\cite{Campos2019_attomolar_DNA} in the LOD. In the following RF and microwave investigation, we focus our attention only on the smallest (and therefore most challenging) concentration, 1 a\textsc{m}. 

\subsection{S-parameter measurements with no gate voltage sweep}

\begin{figure*}[!h]
\centering
\subfloat[]{\includegraphics[width=0.47\linewidth]
{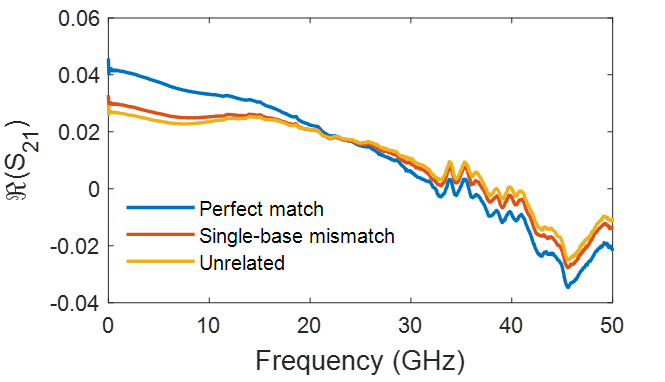}
\label{fig:RealS21}}
\hfill
\subfloat[]{\includegraphics[width=0.47\linewidth]
{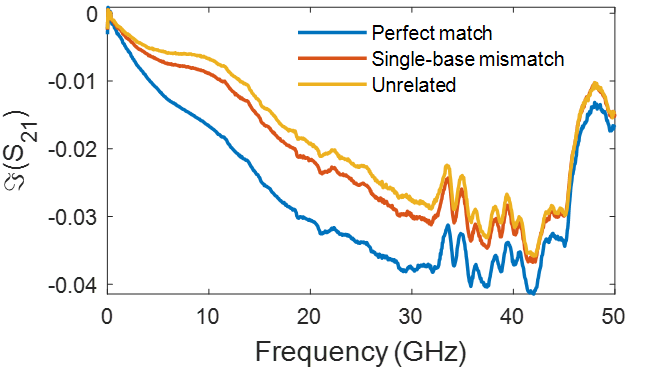}
\label{fig:ImagS21}}
\hfill
\subfloat[]{\includegraphics[width=0.47\linewidth]
{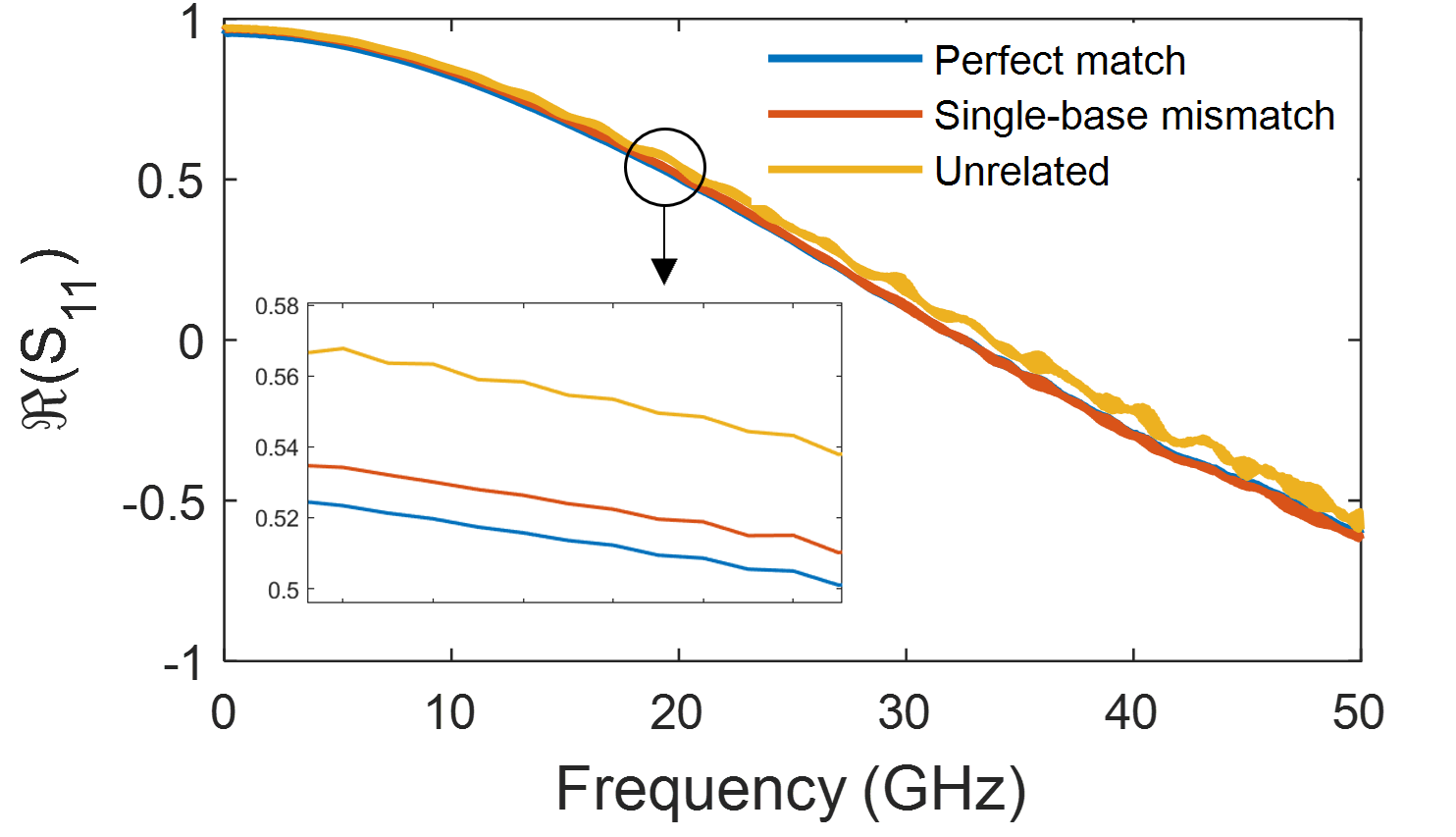}
\label{fig:RealS11}}
\hfill
\subfloat[]{\includegraphics[width=0.47\linewidth]
{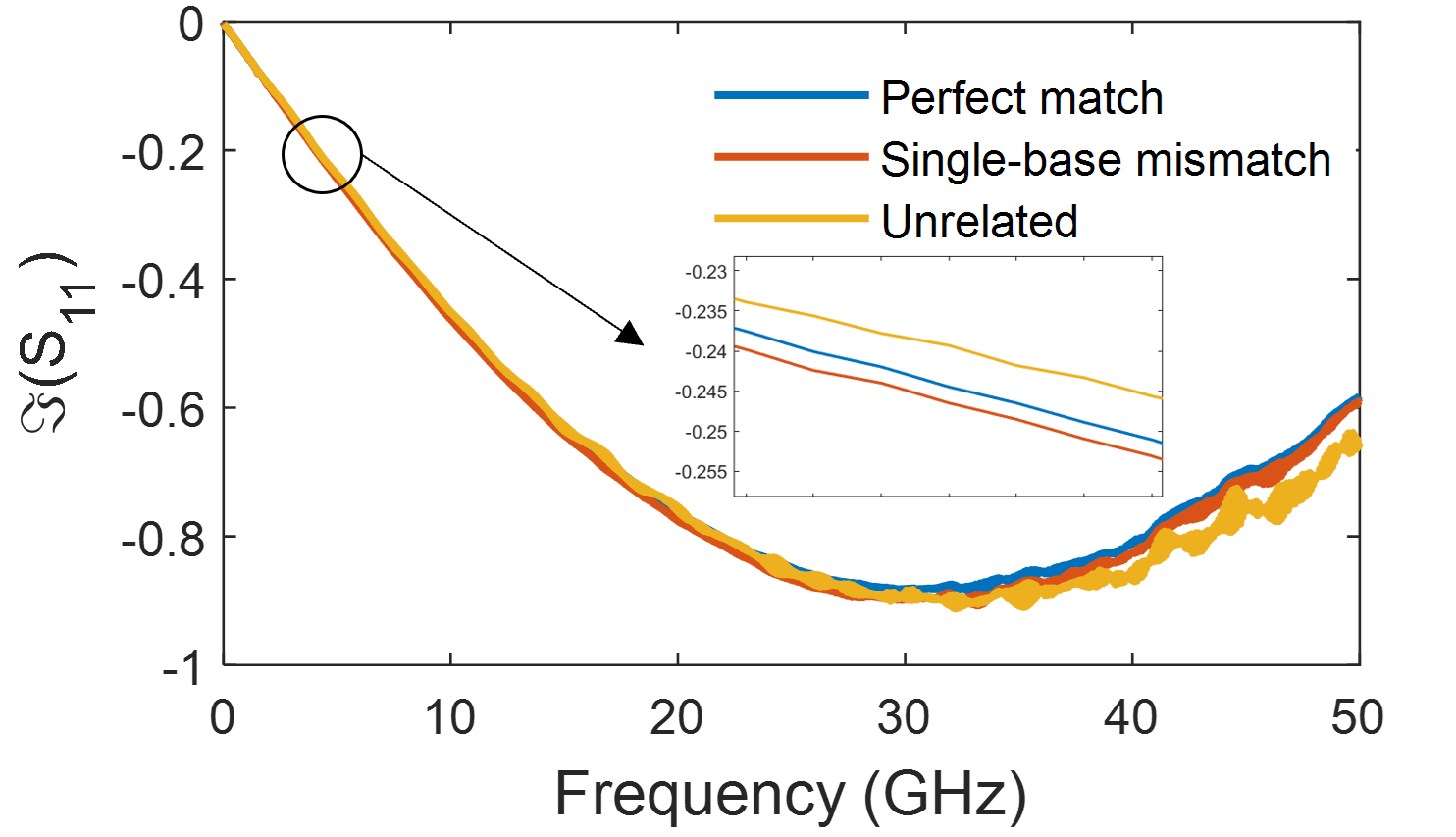}
\label{fig:ImagS11}}
\caption{A representative set of $S-parameter$ curves for a specific device for pmDNA, uDNA, and smDNA. (a) Real part of $S_{21}$. (b) Imaginary part of $S_{21}$. (c) Real part of $S_{11}$. (d) Imaginary part of $S_{11}$. DNA concentration is 1 a\textsc{m}. Gate voltage is \SI{0}{\volt}. The different colours represent different solution types.}
\label{fig:S-parameters2D}
\end{figure*}

The $S-parameter$ curves of pmDNA, smDNA, and uDNA at a concentration of 1 a\textsc{m} of a representative device, with a graphene channel length of $25$ $\mu m$, at a representative gate voltage ($V_{GS}=0$ V), are plotted in \textbf{Figure~\ref{fig:S-parameters2D}}. Clear difference can be observed in the $S_{21}$ parameter (i.e. transmission components), whereas the differences in the $S_{11}$ (i.e. reflection components) are smaller. 

The different frequency dependence trends between different components are of particular interest. For example, the curves of both $\Re(S_{21})$ and $\Im(S_{11})$ of different solutions intersect around 23 GHz, while the other two components present consistent relative magnitudes along the whole frequency range. Also, $\Im(S_{11})$ first decrease and then start to increase at around 31 GHz.
To increase the certainty in the classification of different DNA solutions, we analyse the results at the frequency that indicates the largest measurement differences. The trends imply a certain frequency of around 5 GHz at which the differences in the imaginary components have a high enough SNR, while the trade-off in the differences of the real components is not significant, and keeps the SNR of real components relatively high.
Above $10$ GHz, a decreasing difference in $\Re(S_{21})$ can be seen, while the increase in shifts of the $\Im(S_{21})$ component slows down. By looking at the differences in the components at $5$ GHz with regard to the $0.01\times$ PBS baseline, we can use multiple parameters simultaneously to distinguish smDNA from pmDNA and uDNA.

\begin{figure*}[!h]
\centering
\subfloat[]{\includegraphics[width=0.49\linewidth]
{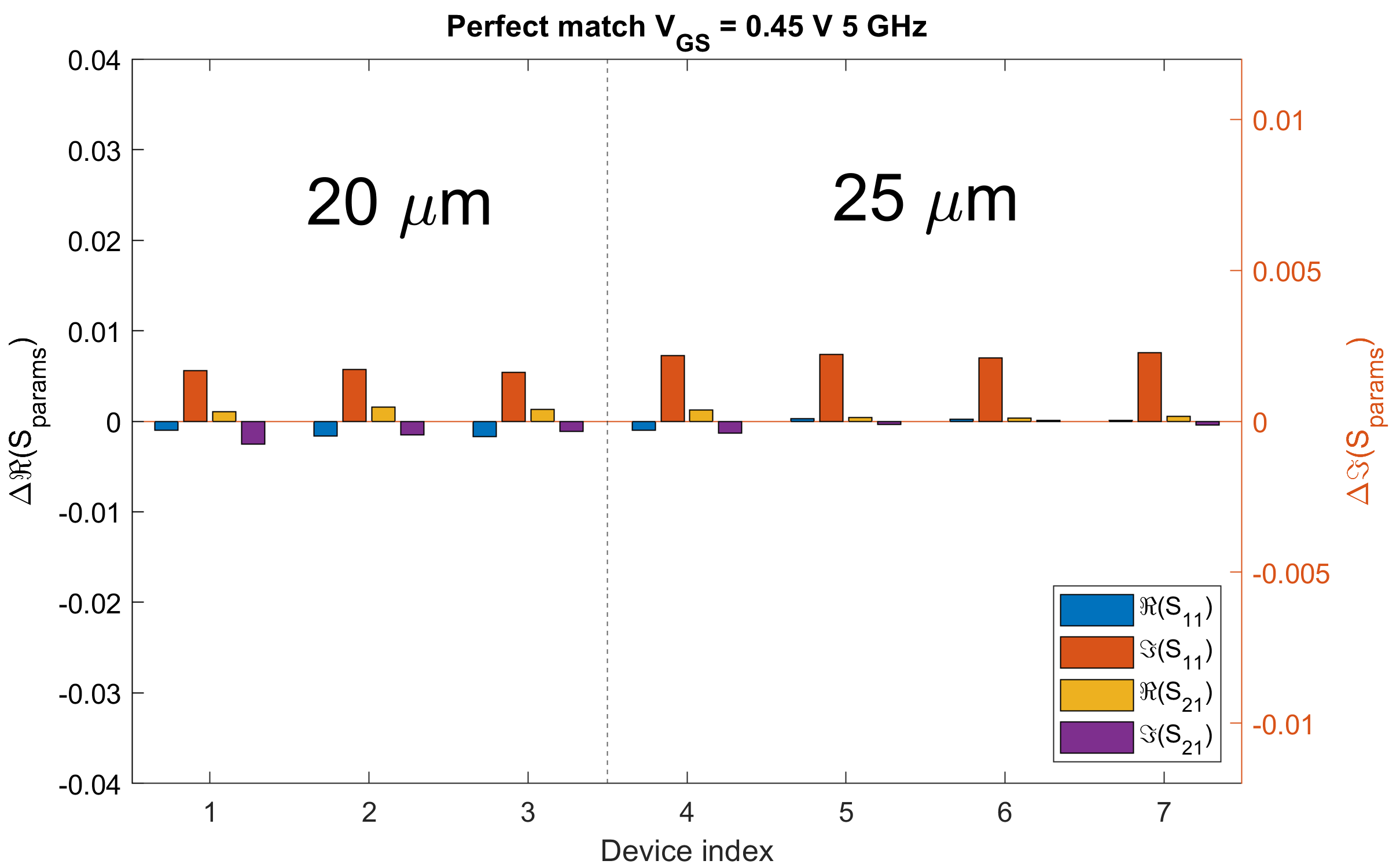}
\label{fig:PM_13devices}}
\hfil
\subfloat[]{\includegraphics[width=0.49\linewidth]
{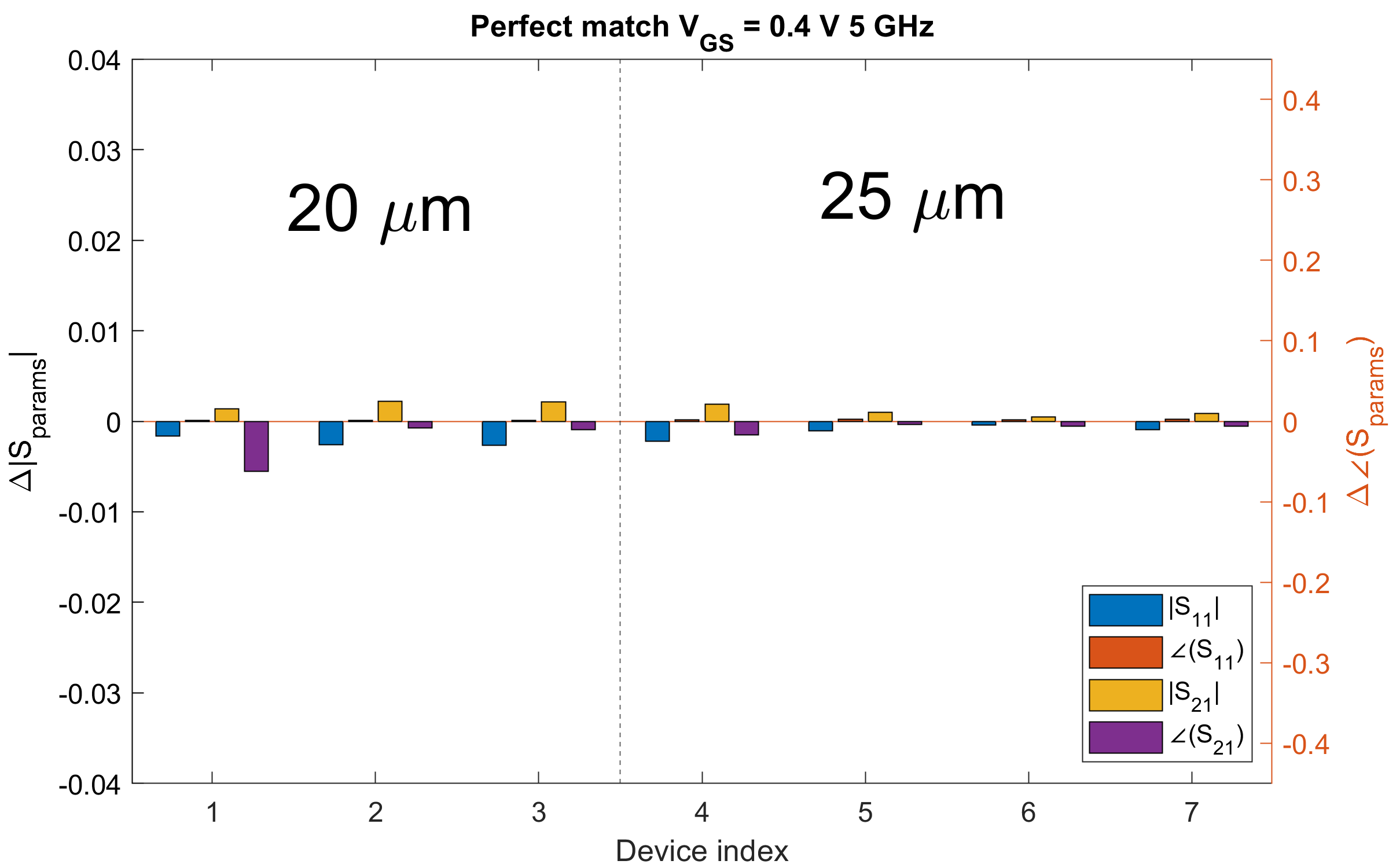}
\label{fig:AMP_PM_13devices}}
\hfil
\subfloat[]{\includegraphics[width=0.49\linewidth]{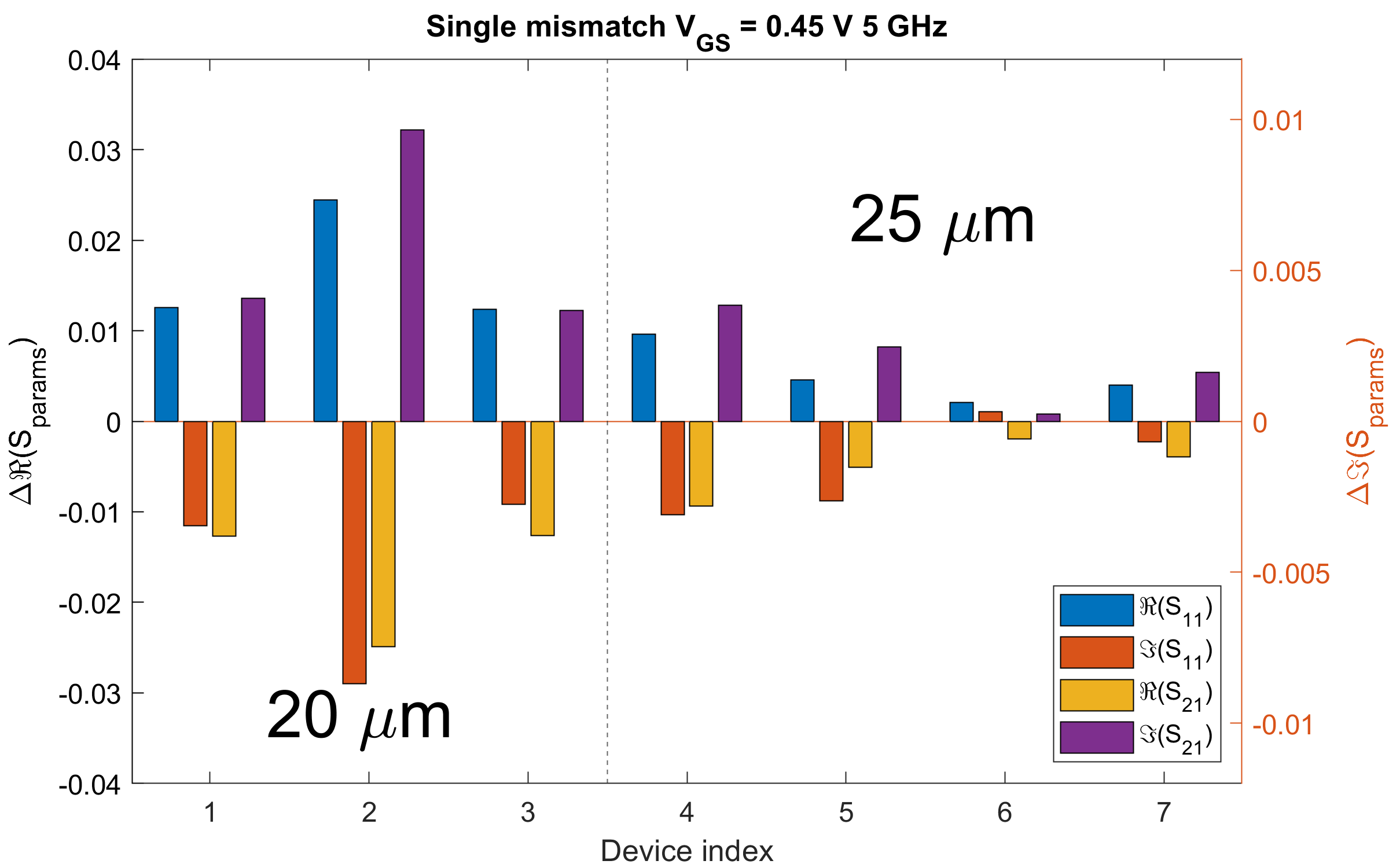}
\label{fig:SM_13devices}}
\hfil
\subfloat[]{\includegraphics[width=0.49\linewidth]{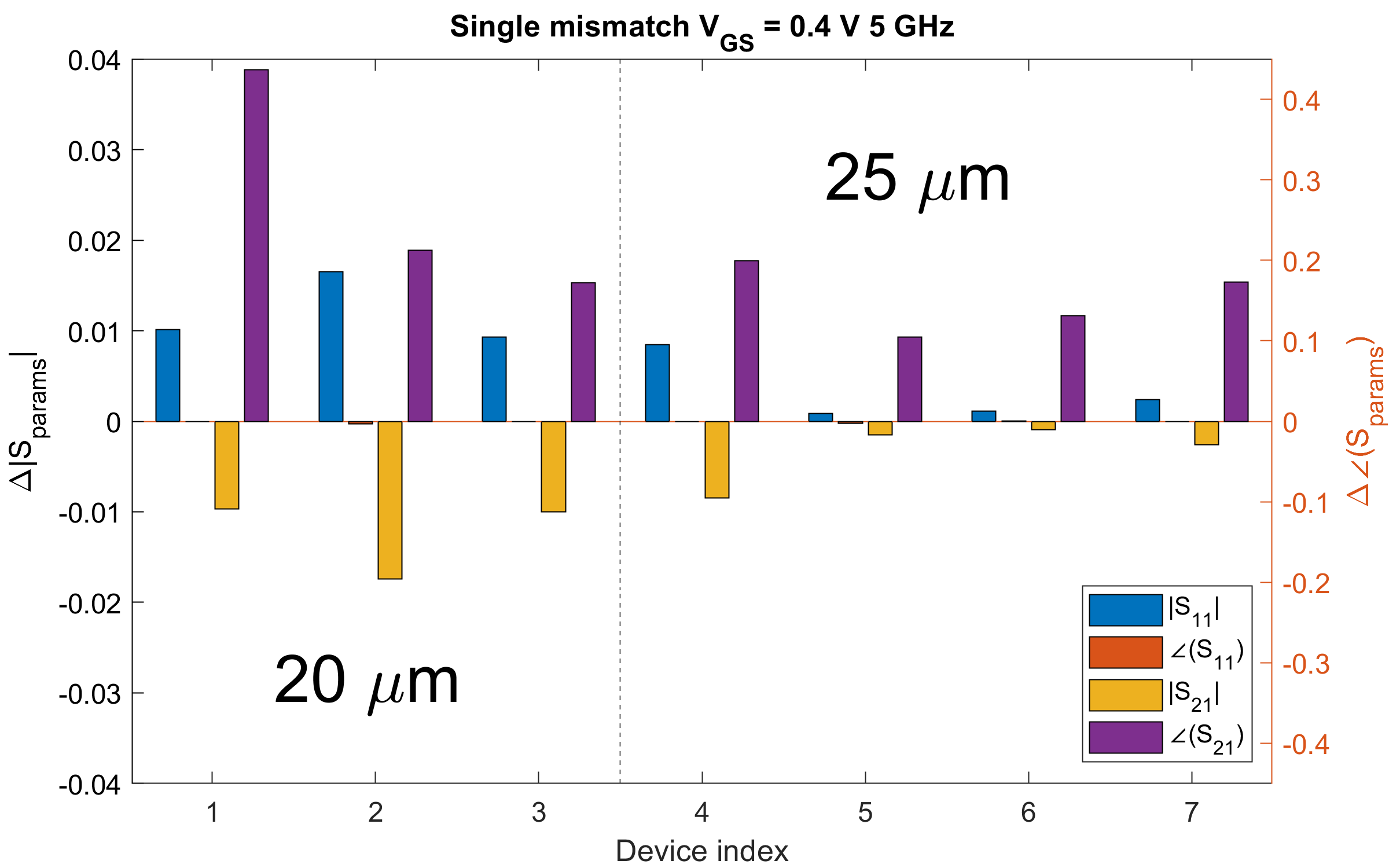}
\label{fig:AMP_SM_13devices}}
\hfil
\subfloat[]{\includegraphics[width=0.49\linewidth]{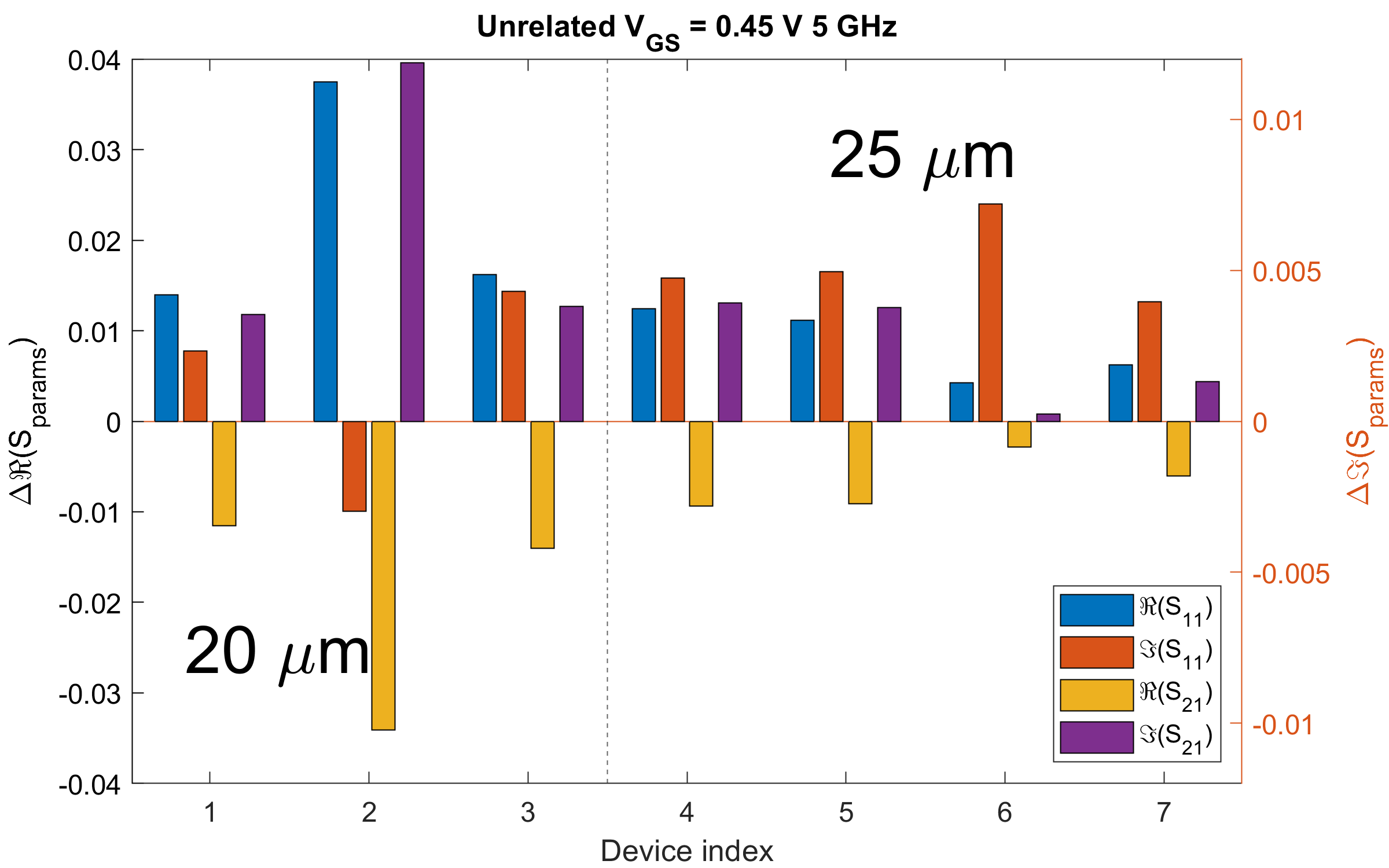}
\label{fig:UM_13devices}}
\hfil
\subfloat[]{\includegraphics[width=0.49\linewidth]{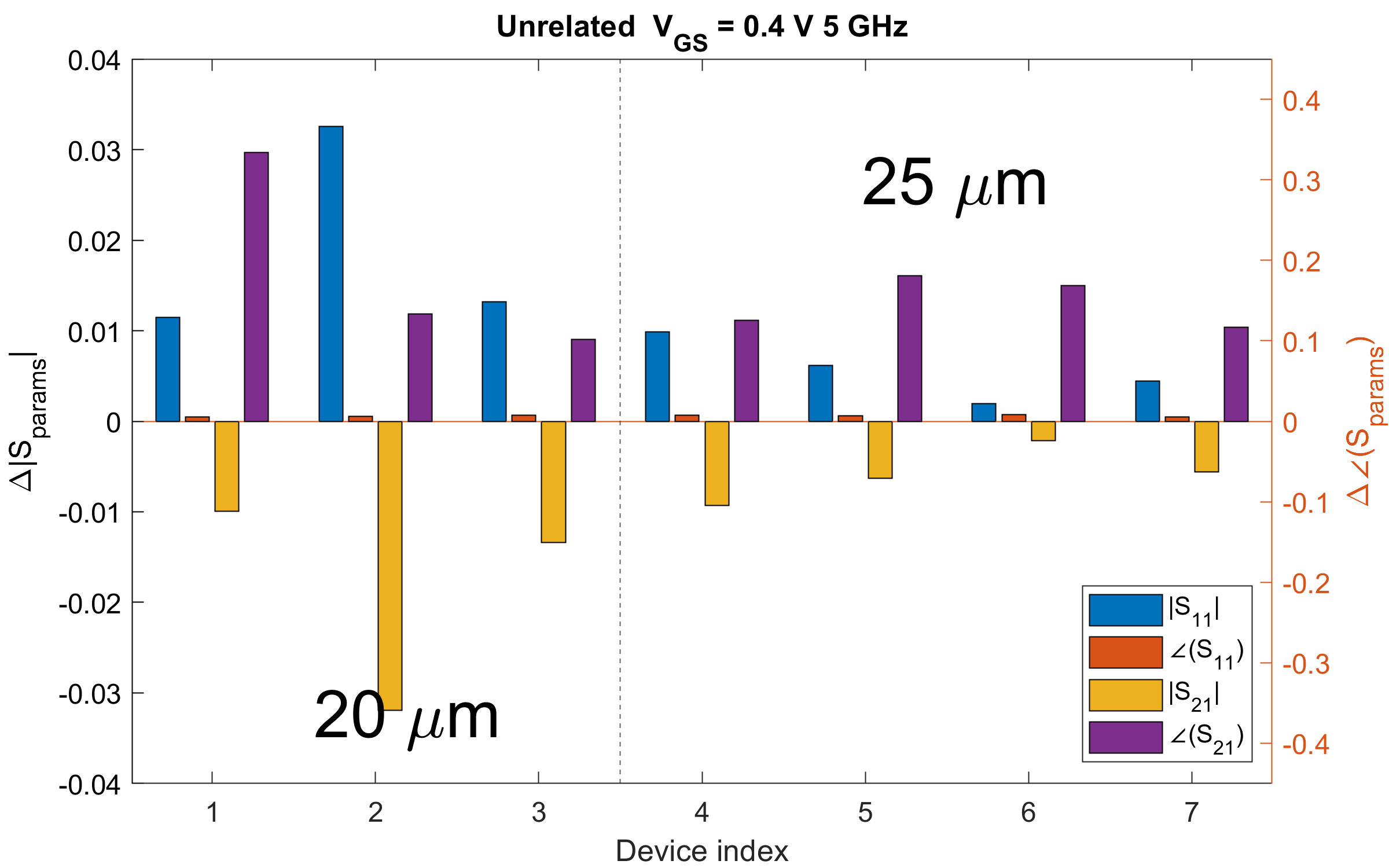}
\label{fig:AMP_UM_13devices}}
\caption{Bar graphs of the differences in the real, imaginary, amplitude, phase of $S_{21}$ and $S_{11}$ between the PBS baseline and pmDNA, smDNA, uDNA, respectively, for devices with a graphene channel length of $20$ $\mu$m or $25$ $\mu$m in the array. Each group of the four bars is the result of an individual device. The frequency is 5 GHz, at which the traces in Figure~\ref{fig:S-parameters2D} present a clear difference between PBS, pmDNA, smDNA and uDNA. DNA concentration is 1 a\textsc{m}. The different colours represent different $S-parameter$ parts. For better visualisation, the y-axis on the left applies to the blue and yellow (Real and magnitude) components, while the y-axis on the right is for the orange and purple (Imaginary and phase) components. The axes were deliberately kept the same for all plots, to better quantitatively visualise the differences.}
\label{fig:S-parameters13devices}
\end{figure*}

To better visualise the differences and validate the reliability of the sensor for the sequence-specific detection of the three types of DNA, we investigate the performance of seven different devices, which have the same design but differ in the length of the graphene channel. \textbf{Figure~\ref{fig:S-parameters13devices}} shows the real, imaginary, amplitude, and phase of the $S-parameter$ differences at 5 GHz between the buffer solution and smDNA, pmDNA, uDNA, respectively, for devices with a channel length of $20$ $\mu m$ or $25$ $\mu m$). We found that longer devices show a consistent response, suggesting that an optimal balance exists between surface area and the losses degrading the SNR. Plots of thirteen devices with channel length varying from $10$ $\mu m$ to $25$ $\mu m$  can be found in the supplementary information. From Figure~\ref{fig:S-parameters13devices} we can see that the differences in the changes of the components are readily apparent for the different sequences. In addition to the differences in the magnitude of the changes, the signs of the changes with respect to the baseline are also different. In particular, different devices present consistent opposite signs of the changes between pmDNA and smDNA. pmDNA always decreases $\Re(S_{11})$, $\Im(S_{21})$, $|S_{11}|$ and $\angle(S_{21})$, and increases $\Im(S_{11})$, $\Re(S_{21})$, $|S_{21}|$ and $\angle(S_{11})$, while smDNA always presents opposite behaviours. The unrelated strand, uDNA, behaves more similarly to smDNA, as expected due to the instability of the hybridization. 

The hybridisation of the pmDNA target with the immobilised probe introduces p-doping and results in shifting the CNP of graphene to more positive values. On the other hand, the produced duplex DNA affects the disorder of carriers and changes the scattering rates of the charge carriers of graphene. uDNA does not bind with probe DNA, and even though the DNA duplexes can still be formed in terms of smDNA, they are unstable~\cite{Campos2019_attomolar_DNA}, leading to reduced electrostatic gating and electron transport in the oligonucleotide and therefore a response similar to uDNA. As a result, different types of DNA modify the graphene's properties and the dielectric properties at the interface differently, leading to differences in the changes in the electrical conductance of graphene and the total capacitance of the channel. These differences impact the transmission and reflection of microwave, which are fully captured by $S-parameters$. 
The consistent differentiation between pmDNA and smDNA results among seven devices showcases the sensor's reliable capability to discriminate pmDNA and smDNA. $\Im(S_{11})$ of uDNA increases in all but one device and can thus be used to distinguish between smDNA and uDNA.

As a result, the differences between the three types of DNA solutions are sufficiently consistent to make logic tables solely on the signs of the changes in S-parameter components with respect to the baseline, as demonstrated in \textbf{Table~\ref{tab:logic_table_ReIm}} and \textbf{Table~\ref{tab:logic_table_AmpPhase}}.

\begin{table}[h!]
\begin{center}
    \begin{tabular}{c|c|c|c|c}
         $\Delta \Re(S_{11})$ & $\Delta \Im(S_{11})$ & $\Delta \Re(S_{21})$ & $\Delta \Im(S_{21})$ & Classification\\
         \hline
         $-$ & $+$ & $+$ & $-$ & pmDNA \\
         $+$ & $-$ & $-$ & $+$ & smDNA \\
         $+$ & $+$ & $-$ & $+$ & uDNA \\
    \end{tabular}
    \caption{The logic table based on the sign of the change of the respective $S$-parameter.}
    \label{tab:logic_table_ReIm}
\end{center}
\end{table}

\begin{table}[h!]
\begin{center}
    \begin{tabular}{c|c|c|c|c}
         $\Delta |S_{11}|$ & $\Delta \angle(S_{11})|$ & $\Delta |S_{21}|$ & $\Delta \angle(S_{21})$ & Classification\\
         \hline
         $-$ & $+$ & $+$ & $-$ & pmDNA \\
         $+$ & $-$ & $-$ & $+$ & smDNA \\
         $+$ & $+$ & $-$ & $+$ & uDNA \\
    \end{tabular}
    \caption{The logic table based on the sign of the change of the respective $S$-parameter.}
    \label{tab:logic_table_AmpPhase}
\end{center}
\end{table}
The above results indicate the feasibility of distinguishing the three types of DNA unambiguously at a specific gate voltage and frequency. The exact behaviour, but not the trend, of each bar chart depends on the gate voltage and frequency at which the difference was calculated. 

\begin{figure*}[htbp]
\centering
\subfloat[]{\includegraphics[width=0.32\linewidth]{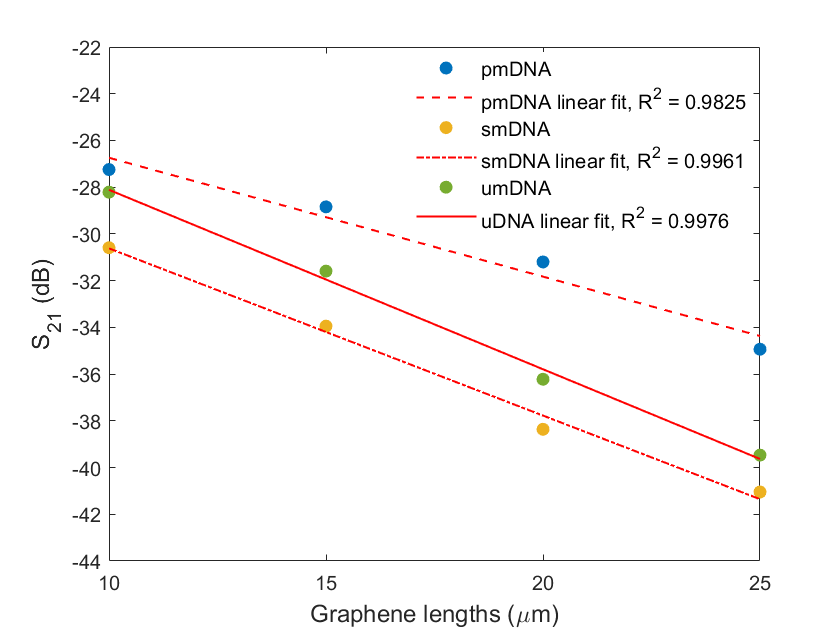}
\label{fig:50MHz}}
\hfil
\subfloat[]{\includegraphics[width=0.32\linewidth]{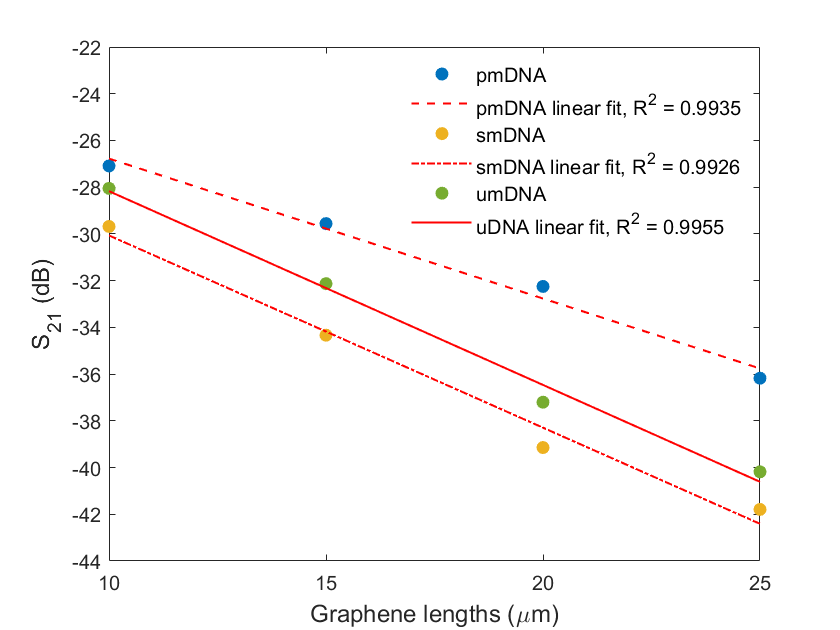}
\label{fig:5GHz}}
\subfloat[]{\includegraphics[width=0.32\linewidth]{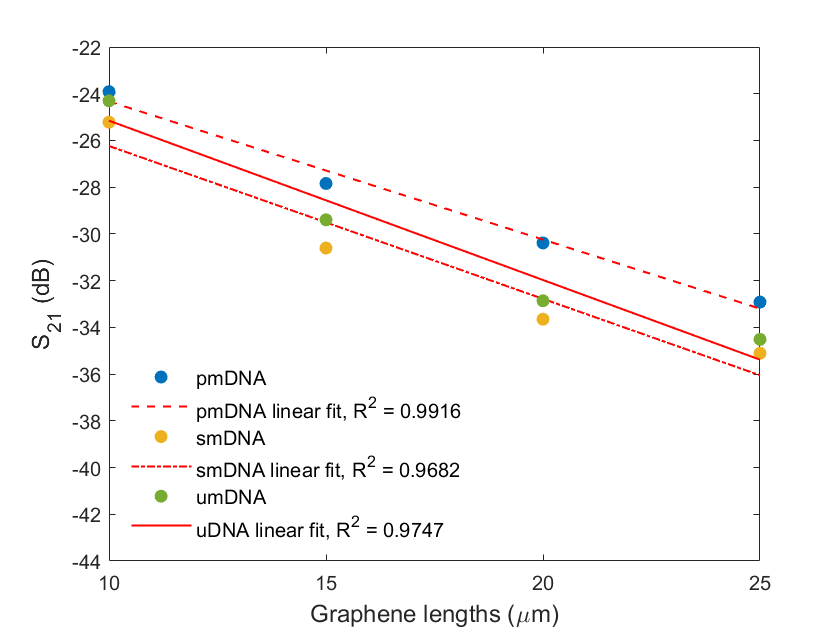}
\label{fig:30GHz}}
\hfil
\caption{Scatter plots and linear-fit curves of $S_{21}$ (dB) against graphene channel lengths for pmDNA, smDNA, and uDNA at (a) 50 MHz, (b) 5 GHz, and (c) 30 GHz. DNA concentration is 1 a\textsc{m}. The good linearity corresponds well with standard wave absorption per the Beer-Lambert law.}
\label{fig:GrapheneLengths}
\end{figure*}

Linear fitting between the graphene channel lengths and the logarithmic (dB) $S_{21}$ at 50 MHz, 5 GHz, and 30 GHz are examined in \textbf{Figure~\ref{fig:GrapheneLengths}}. The logarithmic $S_{21}$ for each graphene length is calculated as the average of the decibel (dB) values of three different devices with the same graphene lengths. 

\subsection{Multidimensional approach}

\begin{figure}[h!]
\centering
\subfloat[]{\includegraphics[width=0.3\linewidth]{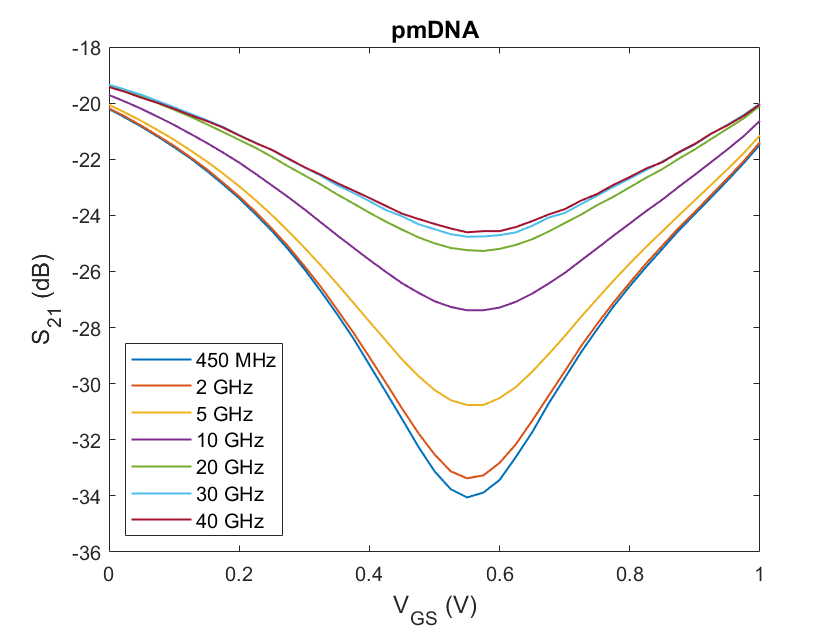}
\label{fig:pmDNA-7frequencies}}
\hfil
\subfloat[]{\includegraphics[width=0.3\linewidth]{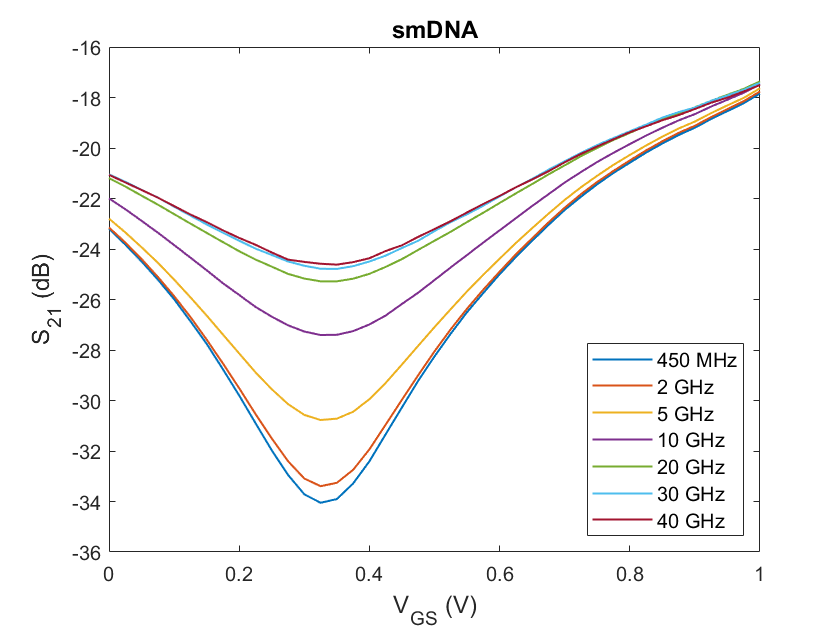}
\label{fig:smDNA-7frequencies}}
\hfil
\subfloat[]
{\includegraphics[width=0.3\linewidth]{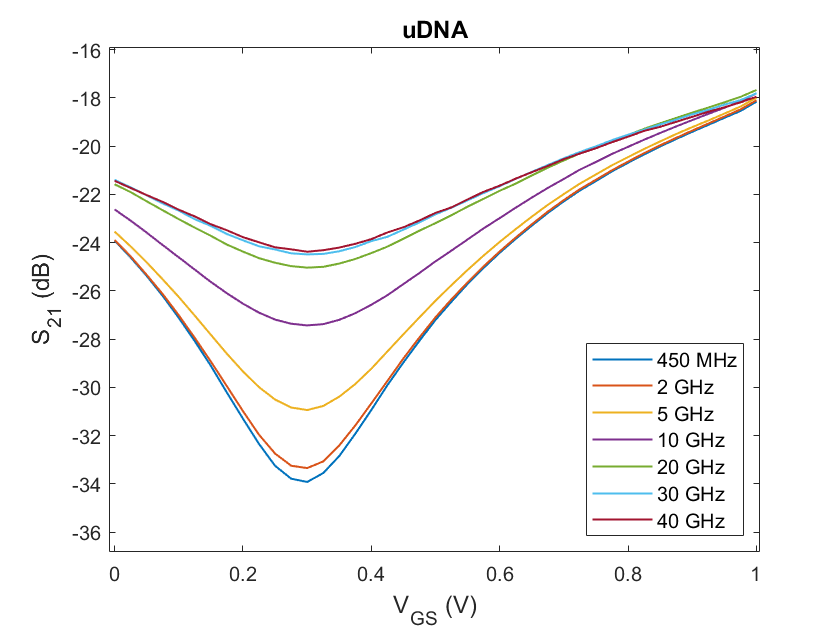}
\label{fig:umDNA-7frequencies}}
\hfil
\subfloat[]{\includegraphics[width=0.31\linewidth]{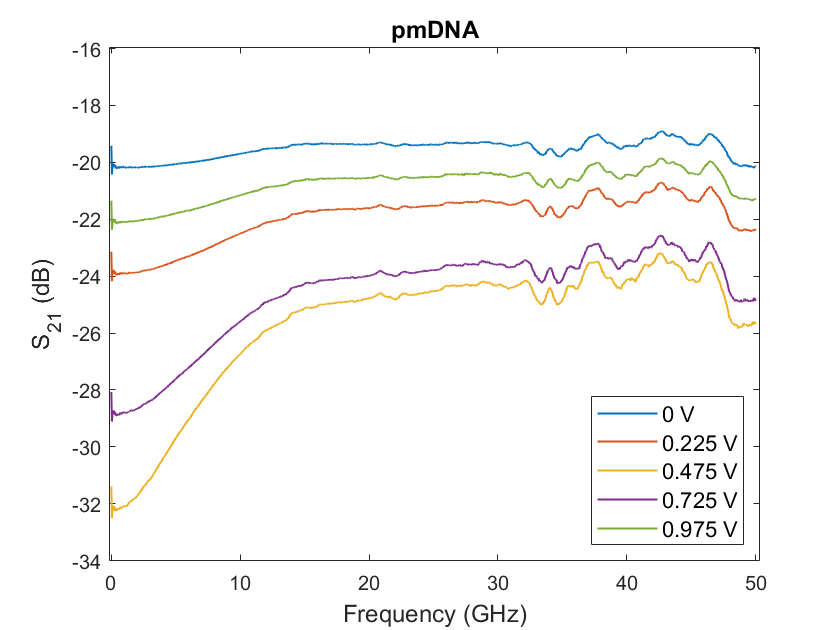}
\label{fig:pmDNA-5volts}}
\hfil
\subfloat[]{\includegraphics[width=0.31\linewidth]{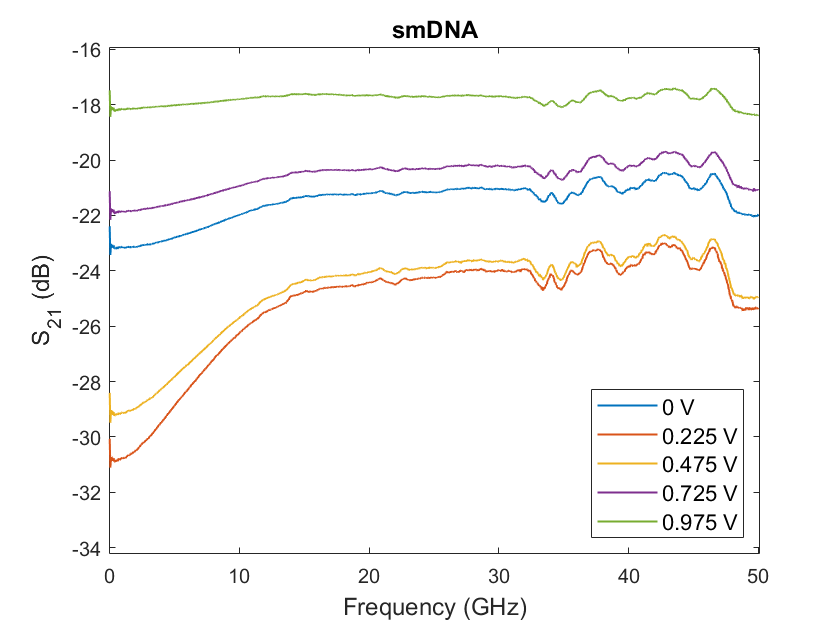}
\label{fig:smDNA-5volts}}
\hfil
\subfloat[]{\includegraphics[width=0.31\linewidth]{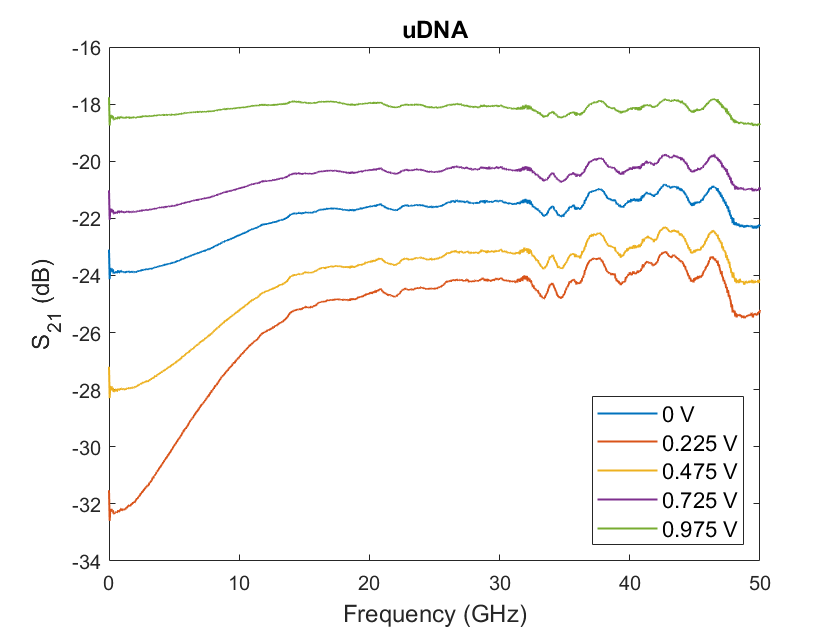}
\label{fig:umDNA-5volts}}
\hfil
\subfloat[]{\includegraphics[width=0.32\linewidth]{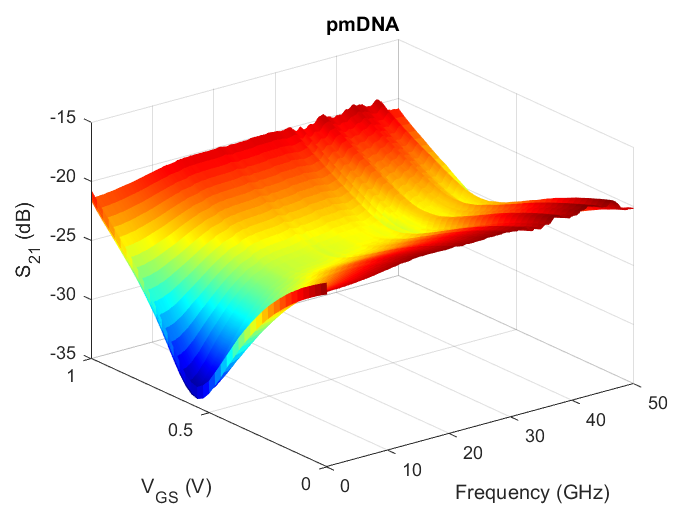}
\label{fig:S21dB-pm}}
\hfil
\subfloat[]{\includegraphics[width=0.32\linewidth]{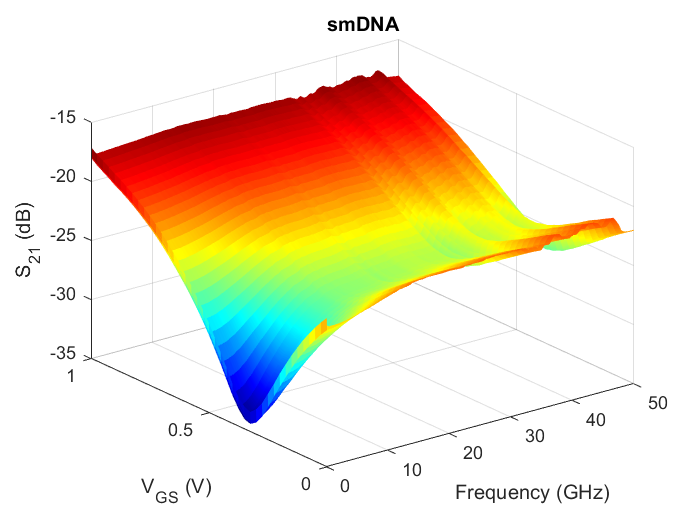}
\label{fig:S21dB-sm}}
\hfil
\subfloat[]{\includegraphics[width=0.32\linewidth]{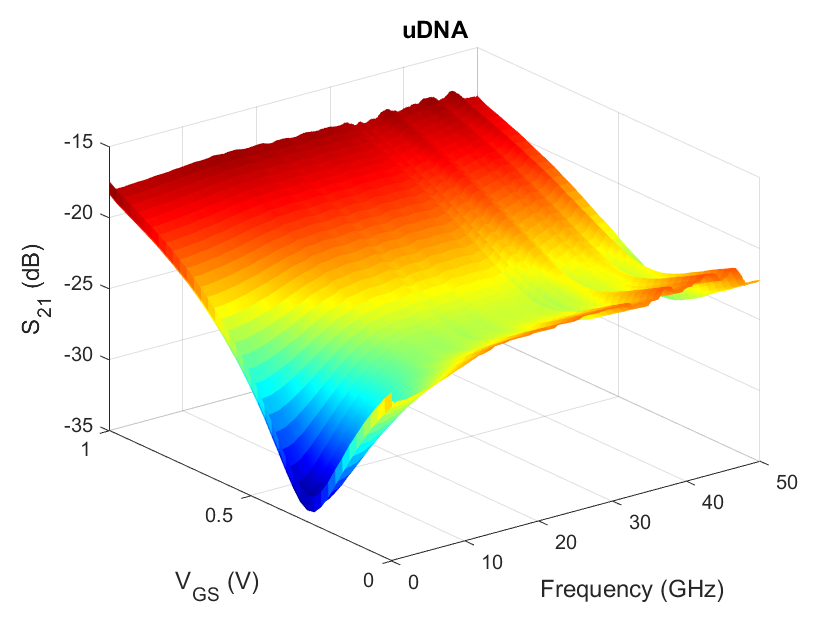}
\label{fig:S21dB-um}}

\caption{$S_{21} \, (dB) - gate \, voltage$ curves at different frequencies for (a) smDNA, (b) pmDNA, and (c) uDNA. $S_{21} \, (dB) - Frequency$ curves at different gate voltages for (d) pmDNA, (e) smDNA, and (f) uDNA. $S_{21} \, (dB) - V_{GS} - Frequency$ 3D plots for (g) pmDNA, (h) smDNA, and (i) uDNA. Variations can be observed as different features at different frequencies and gate voltages on each surface.}

\label{fig:3DS21dB}
\end{figure}

To visualise the effects of different gate voltage ($V_{GS}$) biases, we observe the $S-parameters$ measured with a gate voltage sweep. \textbf{Figure~\ref{fig:pmDNA-7frequencies}}, \textbf{Figure~\ref{fig:smDNA-7frequencies}}, and \textbf{Figure~\ref{fig:umDNA-7frequencies}} exhibit the $S_{21}$ curves against different $V_{GS}$ at seven discrete frequencies for pmDNA, smDNA and uDNA, respectively.
It can be observed that, as expected, the $S_{21}$ (dB) presents a V-shaped trend similar to the DC transfer curves of the device, and the $V_{GS}$ at which the $S_{21}$ (dB) reach minima for the three types of DNA is different due to different binding behaviours. The $S_{21}$ curves against different frequencies at five different gate voltages in \textbf{Figure~\ref{fig:pmDNA-5volts}}, \textbf{Figure~\ref{fig:smDNA-5volts}}, and \textbf{Figure~\ref{fig:umDNA-5volts}} also demonstrate similar trends but different amplitudes at each gate voltage for the three types of DNA. Our design allows independent sweeping of electrolyte gating and wave frequency, which results in $V_{GS} - Frequency$ 3D plots of $S-parameters$. \textbf{Figure~\ref{fig:S21dB-pm}}, \textbf{Figure~\ref{fig:S21dB-sm}} and \textbf{Figure~\ref{fig:S21dB-um}} are the $S_{21} (dB) - V_{GS}-Frequency$ plot at a concentration of 1 a\textsc{m} of one device. The gate voltage sweep and frequency sweep combine to produce rich information that incorporates the variations of the measurements at only one frequency or with a fixed gate voltage bias. As a result, differing from single GFET sensing and single microwave sensing, the response is a combined effect of the changes in graphene conductivity, channel capacitance, and the dielectric properties of the solutions. The resulting variations in the maxima, minima, and gradients of the surfaces offer insights into different hybridisation behaviour of different DNA solutions. Therefore, these observations indicate a good sensitivity of the sensor in the detection the DNA hybridisation at a concentration of 1 a\textsc{m}, and the distinctions between the multidimensional features can be utilised for the discrimination between the three types of DNA. However, upon observation of the multidimensional results in Figure~\ref{fig:3DS21dB}, it is not easy to set a general standard to distinguish the three DNA types, especially in the case between smDNA and uDNA. On the other hand, the information dimensionality is even higher due to the complex values of $S-parameters$, specifically, $[Re(S_{xy})\quad Im(S_{xy})]$ or $[|S_{xy}|\quad \angle S_{xy}]$. As a result, the high-dimensional information and the non-linear behaviours of the sensor encourage the usage of advanced data analysis techniques to obtain quantifiable results and solutions for the classification of the three types of DNA.

\subsection{Classification between single-base mismatch, unrelated, and perfect-match with ML}

ML approaches are suitable for extracting information from rich data. Common ML approaches are investigated to consider the three-class classification task between smDNA, pmDNA, and uDNA at a concentration of 1 a\textsc{M}. The dataset is constructed with the four components of $S_{11}$ and $S_{21}$; $[Re(S_{11})\quad Im(S_{11})]$ and $[Re(S_{21})\quad Im(S_{21})]$ over the whole frequency range and the whole gate voltage range. Specifically, each set of $S-parameter$ components at each gate voltage for each device is a sample. In total, we use $41\times 7$ samples at 41 different electrolytic gate biases from \SI{1}{\volt} to \SI{0}{\volt} and 7 different devices for each DNA class, and each sample has $1001\times 4$ features obtained from the four $S-parameter$ components at 1001 different frequency points from \SI{50}{MHz} to \SI{50}{GHz}. 

To reduce feature dimensions, PCA is utilised, and nine principal components are constructed and kept for the classification. The final dataset has a dimension of $861 \times 9$. This dataset is split into a training set (80\%) and a testing set (20\%), and a validation set is created during training based on different criteria for different algorithms. Popular classification models are implemented on the dataset using MATLAB Classification Learner. The Linear Discriminant Analysis (LDA) has a linear coefficient threshold of 0 and no regularisation. The Support Vector Machine (SVM) has a linear kernel with a box constraint value of 1. The Decision Tree has a maximal number of 100 decision splits. The number of neighbours of the k-nearest neighbours (KNN) is set to 1, with euclidean distance and no distance weight. All the models finish training within 30 s of running on a single core of an Intel i7-12700H CPU. 

\begin{figure}[h]
\centering
\subfloat[Training curves]{\includegraphics[width=0.4\linewidth]{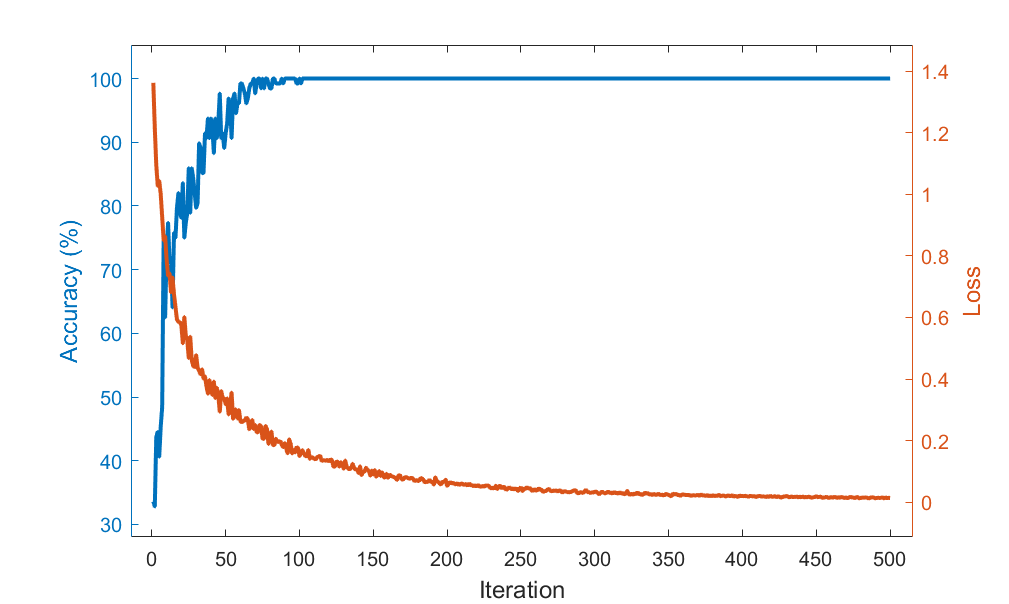}
\label{fig:training}}
\hfil
\subfloat[Validation curves]{\includegraphics[width=0.4\linewidth]{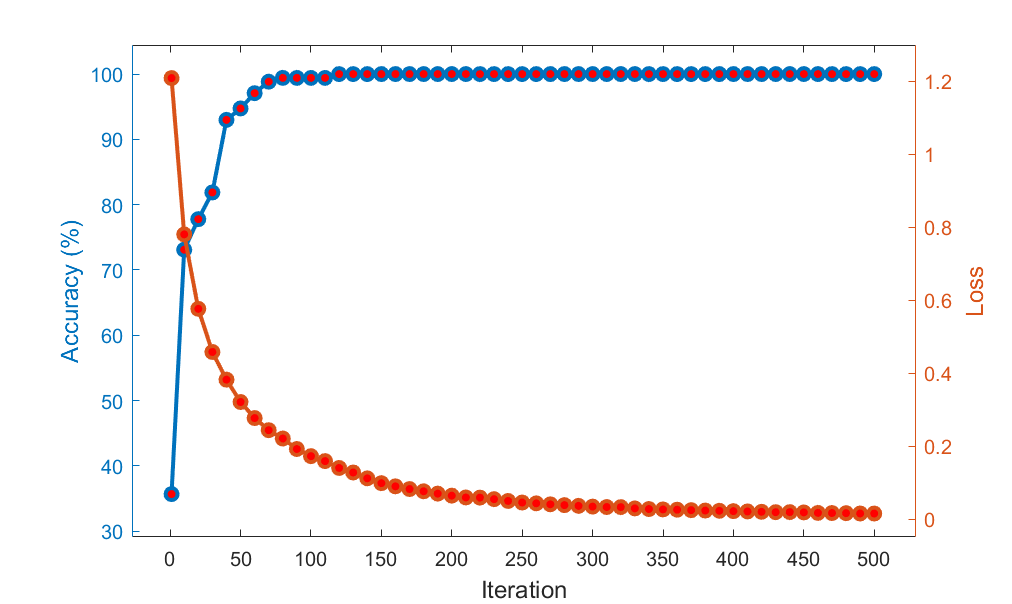}
\label{fig:validation}}
\caption{Accuracy and Loss curves of Training and Validation of ANN. The validation accuracy and loss are evaluated every ten iterations. Both validation and training accuracy increase with iterations and converge to 100\%.}
\label{fgr:ANN}
\end{figure}

\textbf{Table~\ref{tbl:ClassificationResult3}} presents the testing result of different ML models for the three-class classification between pmDNA, smDNA, and uDNA. In addition to the original signal, Gaussian noise is added to the signal to test the robustness of the models. The resulting signal-to-noise ratio (SNR) is 10 dB. We can see that all the models present a testing accuracy of 100\% for the original signal except for the Gaussian Naive Bayes. After adding noise to the signal, the accuracy is not affected much except for the Decision Tree, which indicates the robustness of the models against noise. The Artificial Neural Network (ANN) has an input layer of 4004 and two hidden layers with 20 neurons, each followed by a ReLU activation function layer, and a Softmax (normalised exponential function) layer is added at last to obtain the probability distribution of the outcome for cross-entropy (log) loss calculation. The network is trained with the Stochastic Gradient Descent with Momentum (SGDM) Optimiser to minimise the loss function, with a learning rate of 0.001 and a mini-batch size of 128. A squared magnitude (L2) regularisation term is added with a $lambda$ coefficient of 0.0001. The model is implemented using MATLAB Deep Network Designer. \textbf{Figure~\ref{fgr:ANN}} demonstrates the accuracy and loss curves of training and validation of the ANN model. After around 3 seconds of training, the training accuracy converges to 100\% at around the $75^{th}$ iteration; a similar trend is presented in the validation curve, with a final accuracy of 100\% and loss equal to 0.0026, which attests to the model's low variance.

\begin{table}
\centering
  \caption{ML model classification results}
  \label{tbl:ClassificationResult3}
  \begin{tabular}{lcc}
    \hline
    Algorithms & Accuracy (\%) & Accuracy with noise (\%) \\
    \hline
    LDA  & 100 & 97.66   \\
    SVM & 100 & 99.42 \\
    Tree & 100 & 92.40  \\
Gaussian Naive Bayes & 96.49 & 97.08 \\
    KNN & 100 & 99.42 \\
    ANN & 100 & 99.42 \\
    \hline
  \end{tabular}
\end{table}

\begin{figure*}[htbp]
\centering
\subfloat[]{\includegraphics[width=0.4\linewidth]{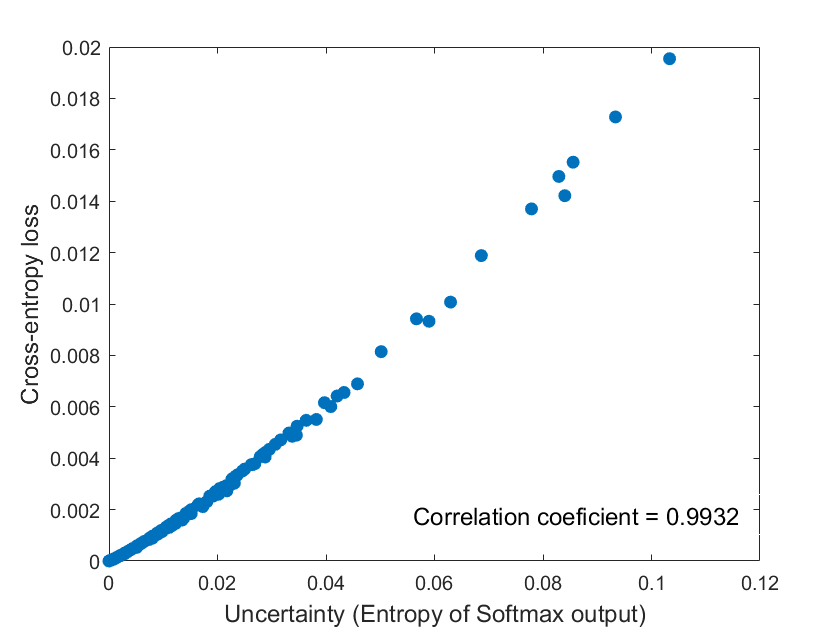}
\label{fig:ANNUncertainty}}
\hfil
\subfloat[]{\includegraphics[width=0.4\linewidth]{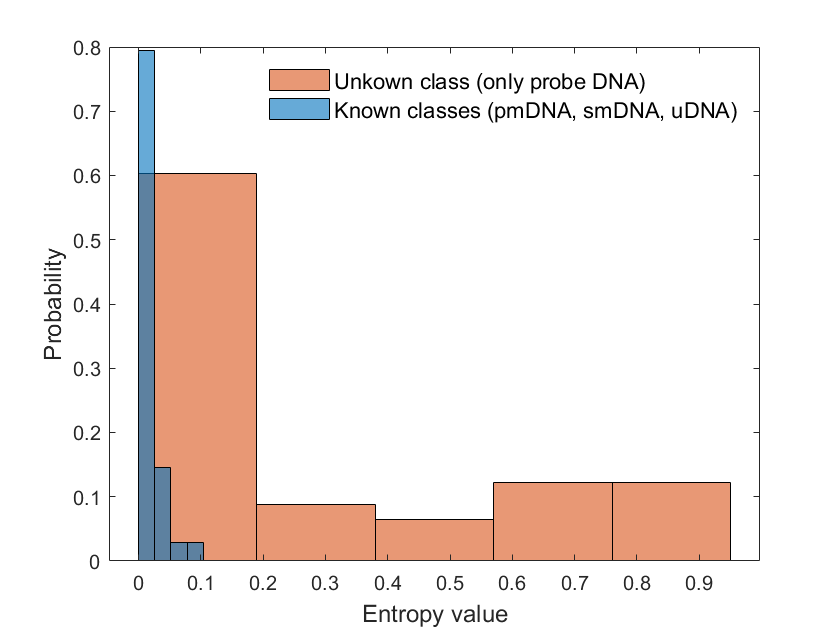}
\label{fig:HistogramOfEntropyForUnknownClass}}
\caption{Uncertainty evaluation of ANN predictions. (a) Correlation scatter plot between the entropy of Softmax output (uncertainty) and the cross-entropy loss at 1 a\textsc{m}. Each dot represents a sample. (b) Histogram of the predictive entropy on testing datasets from known DNA data and unknown probe DNA post-ethanolamine data.}
\label{fgr:ML}
\end{figure*}

The concept of uncertainty has been understudied in the ML context~\cite{AlOsman2021} but is highly relevant for developing measurement methods. Generally, the variance of outputs is used as a measure of uncertainty~\cite{Shirmohammadi2021}. Due to the Softmax probability output, data uncertainty can be quantified for the classification tasks, and the correlation between cross-entropy loss and uncertainty can be calculated to evaluate the uncertainty estimates of the results. The uncertainty estimated by the entropy of Softmax output of the three-class classification results at 1 a\textsc{M} concentration is 0.0159 bit, with a high correlation value of 0.9932. Therefore, the uncertainty in the ML output is low, which implies that the model is certain about its decision, and a high probability is given to the most-likely class. Also, the uncertainty evaluation metric can be considered well-calibrated to predict credibility (uncertainty) approximate to accuracy (error)~\cite{10.1145/1102351.1102430}, and thus the uncertainty estimate can be a useful index for monitoring the performance of the ML model~\cite{10.48550/arxiv.1910.04858}. \textbf{Figure~\ref{fig:ANNUncertainty}} indicates that the two quantities correlate well with each other for the results at 1 a\textsc{m}. However, we observe variations in the uncertainty estimation at different DNA concentrations, for instance, a higher correlation of 0.9932 between uncertainty and loss for the predictions at 1 a\textsc{m} than that of 0.9319 at 1 p\textsc{m}. Therefore, it is worth pursuing uncertainty quantification approaches with a better generalisation that can evaluate both data and model uncertainty~\cite{gawlikowski2021survey}. In addition to adding Gaussian noise to the signal, a method based on uncertainty evaluation is also conducted to examine the performance of the ANN model. A robust model should output high predictive uncertainty if the input is unknown~\cite{NIPS2017_9ef2ed4b}. The histogram of the Softmax entropy uncertainty score on test examples from known classes (pmDNA, smDNA, and uDNA) and the unknown class (only probe DNA) is shown in \textbf{Figure~\ref{fig:HistogramOfEntropyForUnknownClass}}. We can observe a clear increase in the output entropy when the model handles the unknown data,  which validates the credible predictions for the known classes of DNA. The ability of ML models to distinguish out-of-distribution examples is critical for applications in practical systems. 

In summary, by using ML, high accuracy and reliability in the detection of single-base mismatch are achieved based on the multidimensional broadband RF signal measured with the graphene microwave sensor. The models are applicable to the data measured with different gate voltage biases from 0 V to 1 V. Therefore, whereas the logic table approach only takes into account a specific gate voltage and frequency combination, the machine learning-assisted multidimensional approach considers the full dataset, i.e. all the gate voltages as the frequencies measured, thus providing a robust classification between the three sequences considered. Note that experimental data does not require precise and complicated calibrations and de-embedding before being fed to ML models for training and predictions, which reduces experimental work and artefacts and thus increases efficiency and accuracy. Also, ML can extract features from the measurement data containing outliers and noise. Moreover, the short training time of ML models and the advantage of predicting new data using the trained models instantly make ML well-suited for real-time applications.

\section{Conclusions}

We presented a novel DNA sensor consisting of electrochemically-gated graphene coplanar waveguides coupled with a microfluidic channel. By immobilising probe DNA sequences on the graphene surface and exposing the sensor to either perfect-match DNA, single-base mismatch DNA, or unrelated DNA, we achieve consistent and reproducible identification of single-base mismatch at DNA concentrations as low as 1 a\textsc{m}. The possibility of independently controlling gate voltage and frequency sweeps leads to multidimensional datasets, which are analysed using machine learning methods. The machine learning results possess an improved generalisation for the DNA discrimination in comparison with using the data at only one gate voltage bias and one frequency. A classification accuracy of 100\% between single-base mismatch, unrelated, and perfect match is achieved, as well as a classification accuracy \textgreater{} 97\% in the presence of simulated Gaussian noise. The device concept presented here goes beyond DNA sensing, paving the way for a generally-applicable approach to biosensing, where chemical field-effect sensing and microwave impedance spectroscopy are combined in a single platform.

\section{Methods}

\subsection{Graphene synthesis and transfer}
Graphene layers are grown via Chemical Vapour Deposition (CVD) on Cu foil, following the protocol developed by Burton et al~\cite{Burton2019}. The foil is heated to a high temperature, and hydrogen and methane gases are introduced. The copper foil acts as both the catalyst and the substrate for the deposition. Graphene is transferred onto high-resistivity Si substrates ($\rho >10,000$ $\Omega\cdot$cm) covered in 300 nm-thick thermally-grown oxide by using a sacrificial layer of polymethyl methacrylate (PMMA) 495K (8\% in anisole, MicroChem),  deposited onto the graphene-coated Cu foil via spin-coating. After exposing the bottom side of the Cu foil to oxygen plasma ($10$ W, $60$ s), the PMMA/graphene/Cu stack is then floated on a $0.5$ $m\%$ solution of \ce{(NH4)2S2O8} overnight. This etches away the metal layer, leaving the graphene-PMMA stack floating on the surface. This stack is then picked up and re-floated on top of DI water to remove metal and etchant residue. To ensure good adhesion between the graphene and the substrate as well as reduce the interaction with the Si-O-H groups on the surface, any trapped water and the graphene, the wafer was primed using Hexamethyldisilazane (HMDS). The wafer is then used to lift the graphene-PMMA stack of the DI water surface and left to dry naturally overnight. The sacrificial PMMA is removed by two consecutive acetone baths and an isopropanol bath. After the PMMA removal, the graphene on the Si substrate is heated to $150$ \si{\celsius} for $5$ minutes to drive out any residual moisture trapped between the substrate and graphene.

\subsection{Design and fabrication}
The largest part of the CPW, which feeds the signal to the graphene channel, is designed to match Ground-Signal-Ground (GSG) probes with $150$ $\mu$m pitch per manufacturer specification~\cite{cascade_layout}. For better adhesion of the microfluidic polydimethylsiloxane (PDMS) channel, the metal part of the waveguide is extended to maximise the contact area between PDMA and the substrate. The GFET is incorporated by tapering the central trace of the waveguide and adjusting the separation between the G and S conductors so that the characteristic impedance is close to $50$ \si{\ohm}, matching the rest of the measurement system. Most graphene FET devices, including state-of-the-art sensors, use a graphene-on-top design. The advantage of our approach is the ability to pre-treat the graphene-metal contact area prior to deposition to improve the contact resistance, especially important for RF devices~\cite{Cusati2017,Anzi2018}. The increased length of the contact area beyond that of typical DC devices also increases the capacitive coupling between the metal leads and the graphene channel, resulting in lower insertion loss in comparison with previous efforts~\cite{gubeljak_Memea2022}. The dried graphene is patterned using direct-write laser lithography, and the excess material is etched by reactive ion etching using $3$ W oxygen plasma for $30$ seconds. The metal layer is patterned with a long overlap between the exposed graphene. Directly prior to metal deposition, we expose the contact area using a $0.5$ W argon plasma for 20 seconds to improve our contact resistance. We use $5$ nm Cr as the adhesion layer and deposit a further $100$ nm of gold using an e-beam evaporator followed by liftoff in acetone. The microfluidic channel was made using PDMS from an SU-8 epoxy mould and mounted to the patterned CPW structures. To connect the device assembly to the microfluidic controller, feedlines were introduced and the entry points were sealed using liquid PDMS. To bond the PDMS to the wafer better and cure the fluid inlet seals, we cured the entire assembly at $80$ \si{\celsius} overnight.

\subsection{Functionalisation}
We use the microfluidic assembly to passivate and functionalise the metal and graphene areas of the CPWs, respectively. First, the Au is passivated using a solution of 1-dodecanethiol, as oligonucleotides have an affinity for Au. Then, the linker molecule, PBASE, is introduced and left for 2 hours to interact with the graphene, after which the channel is rinsed using methanol. This allows us to immobilise a probe single DNA strand consisting of 20 bases (P20 - GAGTTGCTACAGACCTTCGT) on the graphene surface using the added amine (\ce{NH2}-) group at the 5' end by exposing the graphene channel to a slow flow ($0.5$ $\mu$L/min) of a 10 $\mu$M dispersion of P20 in a 1\% phosphate buffer solution ($0.01\times$ PBS) for 12 hours. To prevent any unreacted PBASE sites from reacting with our target oligonucleotides (pmDNA,smDNA and uDNA), we passivate them using a $0.1$ M ethanolamine solution in $0.01\times$ PBS for 30 min. The post-ethanolamine measurement thus forms our baseline for comparing all other data. The pmDNA sequence is ACGAAGGTCTGTAGCAACTC, the smDNA sequence CCGAAGGTCTGTAGCAACTC, and the uDNA sequence TAGTATAGTTTGGATGTACA, synthesised by Sigma Aldrich. 

\subsection{Measurements}
For DC and RF characterisation, we mount the sample on a Cascade Microtech Summit 12000B semi-automatic probe station. An Agilent PNA-X 5245 vector network analyser (VNA) is connected to a pair of Cascade Microtech i50 GSG 150 Infinity probes, which we calibrate using a manufacturer-provided impedance standard substrate (ISS 101-190C) via the mTRL method. For DC measurements and electrolytic gating, we use an Agilent B1500A Parameter Analyser, connected to the RF probes using a biasing network and to the gate electrode using a Cascade Microtech DCP100 probe, as shown in Figure~\ref{cpw_sketch} (d) \& (e). We used an RF power of $-12$ dBm and an IF of $1$~kHz, with $1001$ points collected between $50$ MHz and $50$ GHz. The fluid flow is controlled by an ElveFlow OB1+ Mk3 microfluidic controller. Each analyte, i.e. pmDNA, smDNA or uDNA is introduced by setting the flow to $1$ $\mu$L/min and reacting for $60$ minutes, after which the channel is rinsed with $0.01\times$ PBS for $10$ minutes. All measurements are performed with $0.01\times$ PBS flowing into the channel at $1$ $\mu$L/min. We started by measuring the perfect match DNA dispersion, followed by a $0.1$ M NaOH solution and buffer rinse, to recover the pristine probe DNA sites, as per~\cite{Xu2017_realtime_DNA}. The matching concentration of smDNA was then introduced and measured in the same manner, followed by another NaOH rinse and uDNA. As GFETs often exhibit hysteresis, we swept the gate voltage from $0$ V to $1$ V, with a step size of $25$ mV and stabilisation delay of 1 second, $10$ times to stabilise the transfer curves, after which the $S$-parameters are measured. To accurately and repeatably measure the array of devices, we developed automated control scripts to perform DC measurements, synchronise $S$-parameter measurements with the applied electrolytic gate bias and characterise the entire wafer.

\medskip
\textbf{Acknowledgements} \par 
This work is supported by EPSRC grant EP/L016087/1. A.L. acknowledges Gos Micklem for useful discussions. 

\medskip

\bibliographystyle{ieeetr}
\bibliography{bibliography}

\end{document}


\title{Supplementary information}

\maketitle
\vspace{-1cm}
$^\dagger$ These authors contributed equally to this work. *Corresponding author: \href{mailto:a.lombardo@ucl.ac.uk}{a.lombardo@ucl.ac.uk}

\newpage
\section{S-parameters}
\subsection{S-parameter plots of devices with channel length from 10 to 25 $\mu m$}

\begin{figure*}[!h]
\centering
\subfloat[]{\includegraphics[width=0.40\linewidth]
{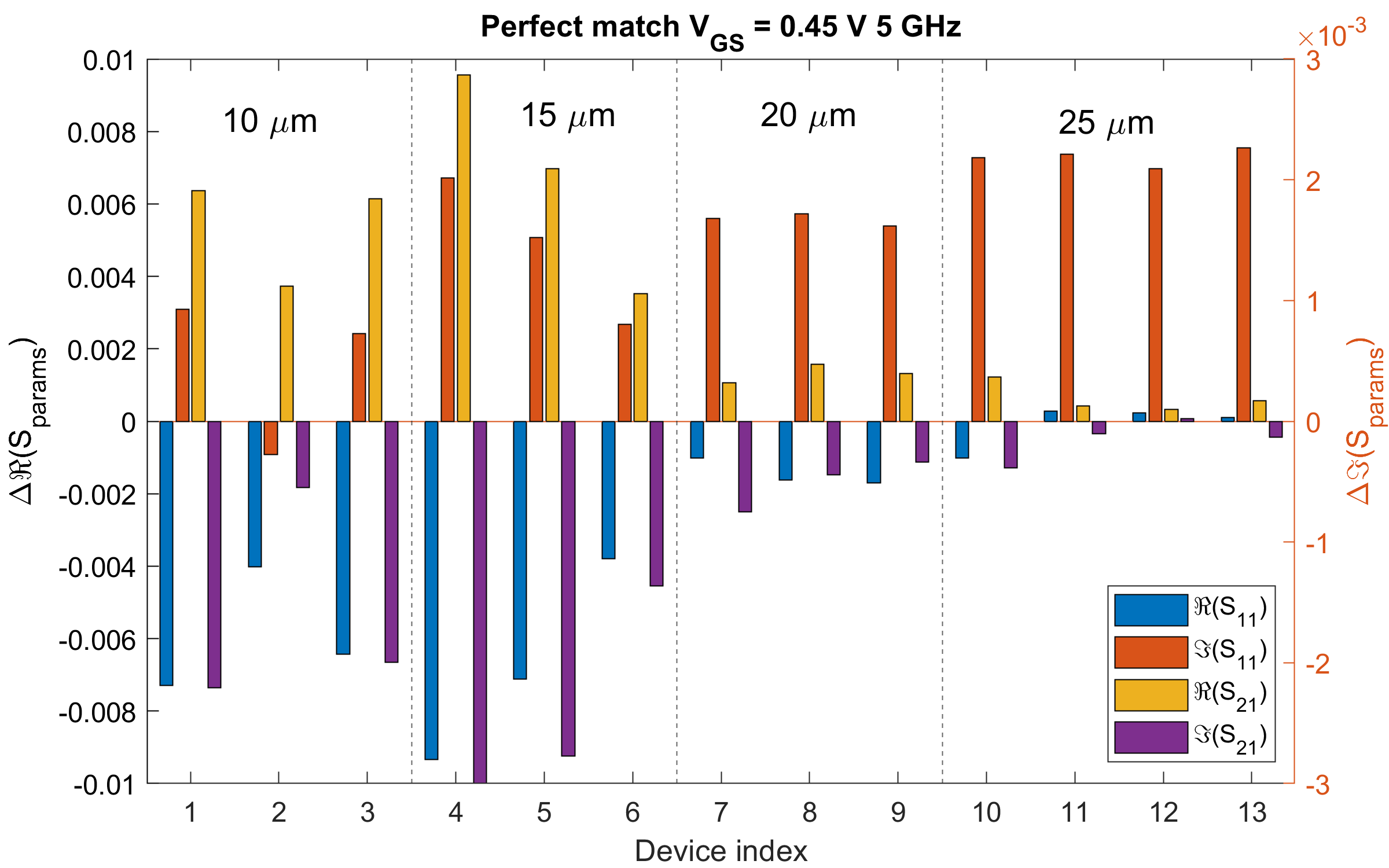}
\label{fig:PM_13devices}}
\hfil
\subfloat[]{\includegraphics[width=0.40\linewidth]
{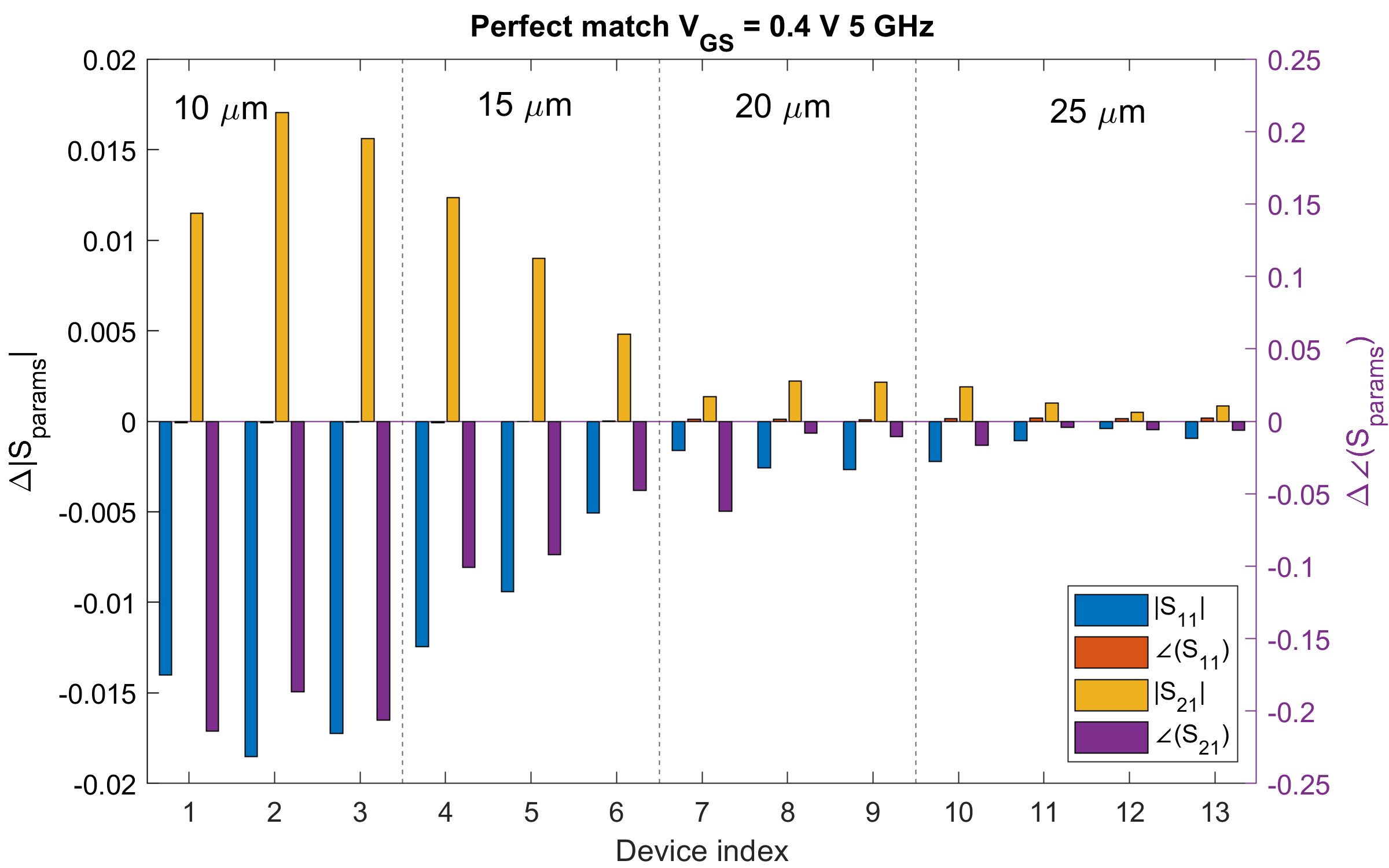}
\label{fig:AMP_PM_13devices}}
\hfil
\subfloat[]{\includegraphics[width=0.40\linewidth]{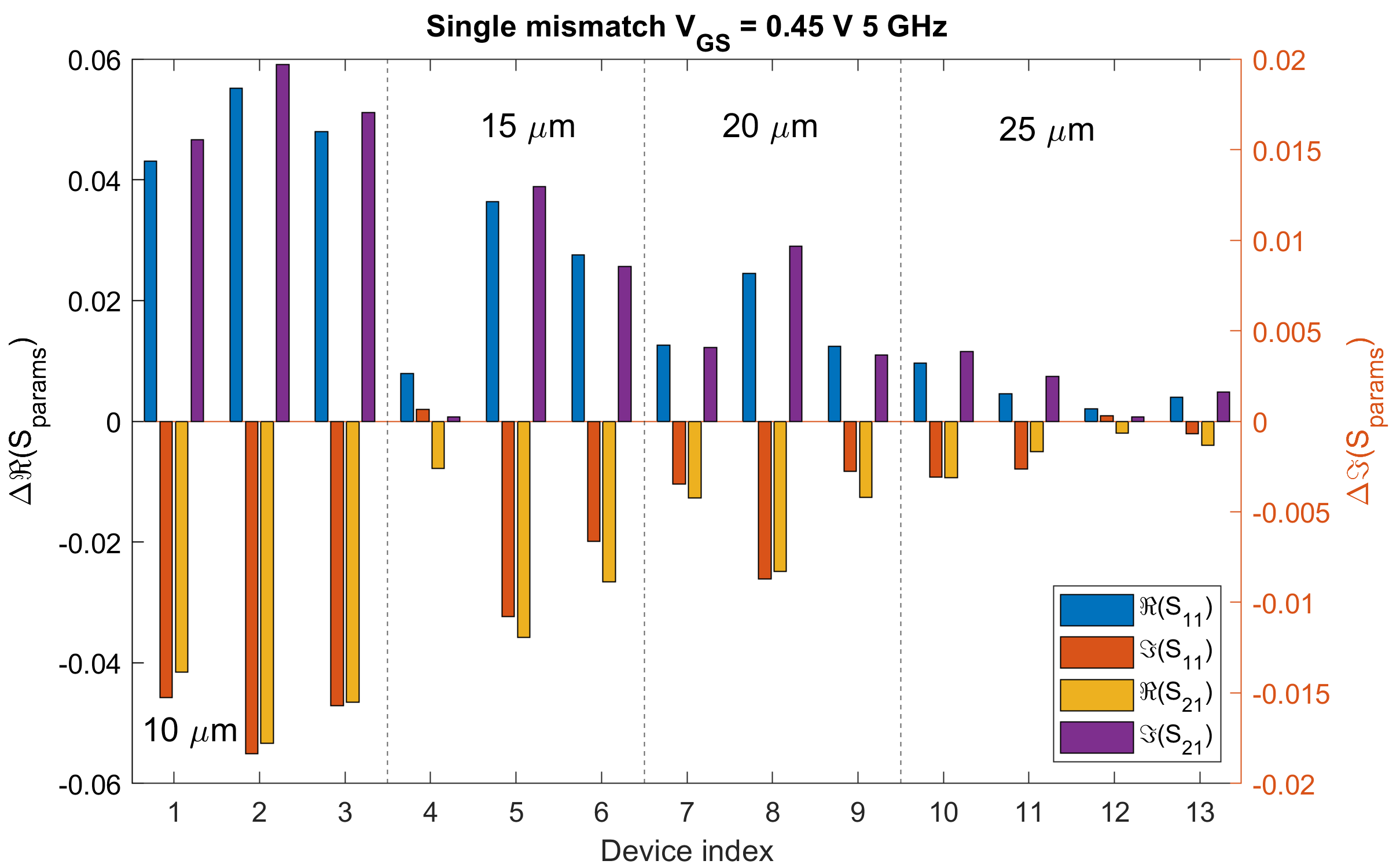}
\label{fig:SM_13devices}}
\hfil
\subfloat[]{\includegraphics[width=0.40\linewidth]{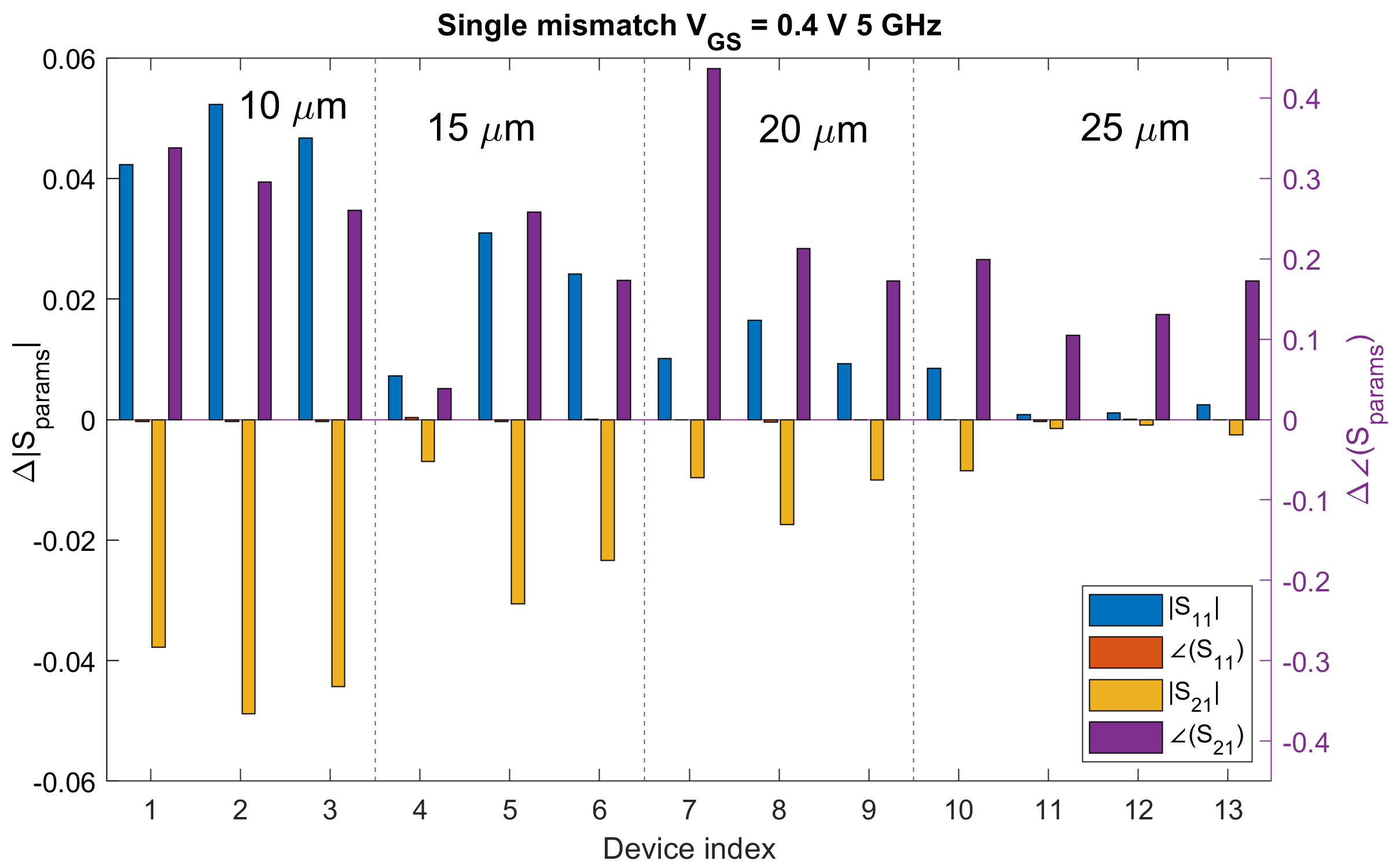}
\label{fig:AMP_SM_13devices}}
\hfil
\subfloat[]{\includegraphics[width=0.40\linewidth]{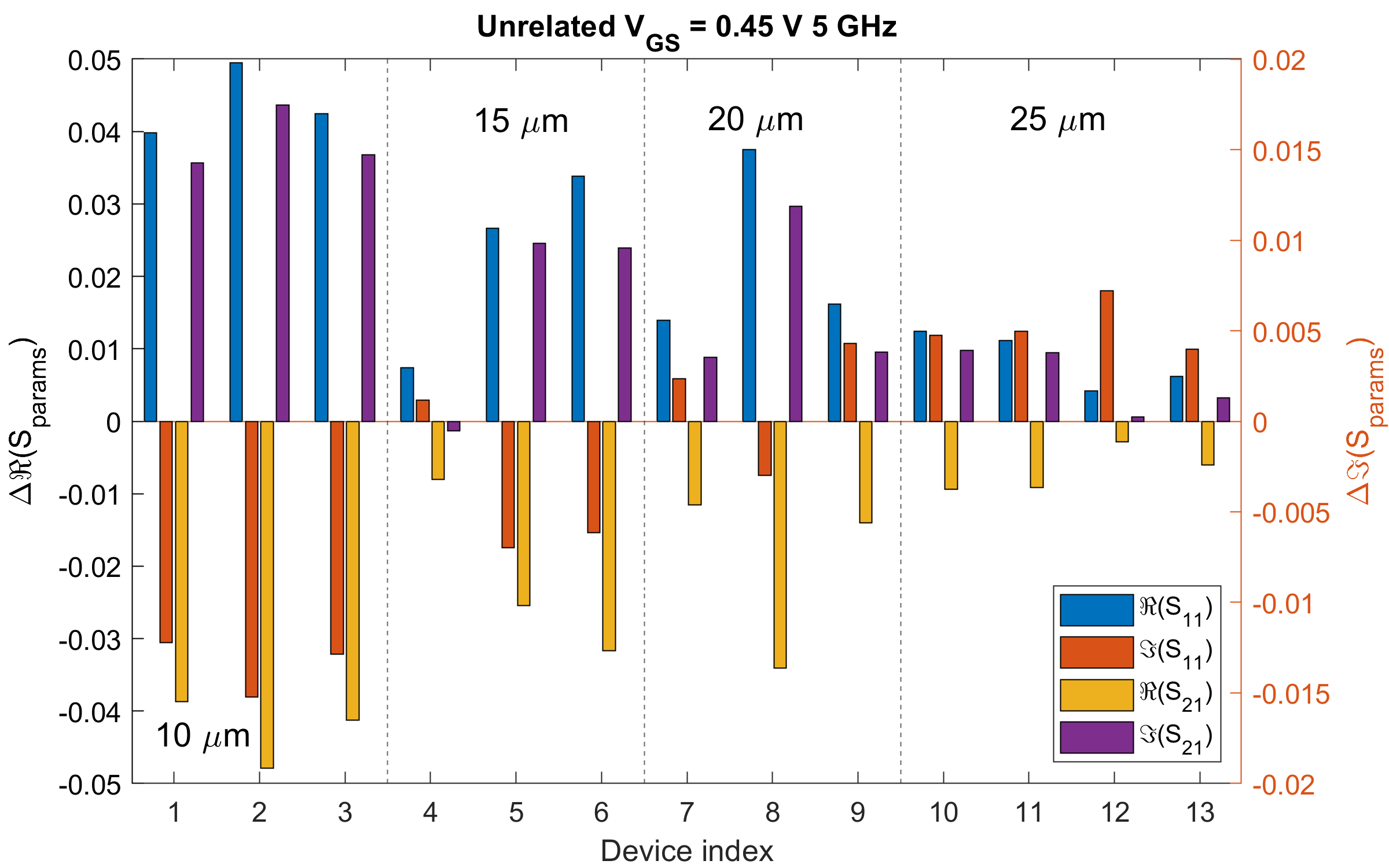}
\label{fig:UM_13devices}}
\hfil
\subfloat[]{\includegraphics[width=0.40\linewidth]{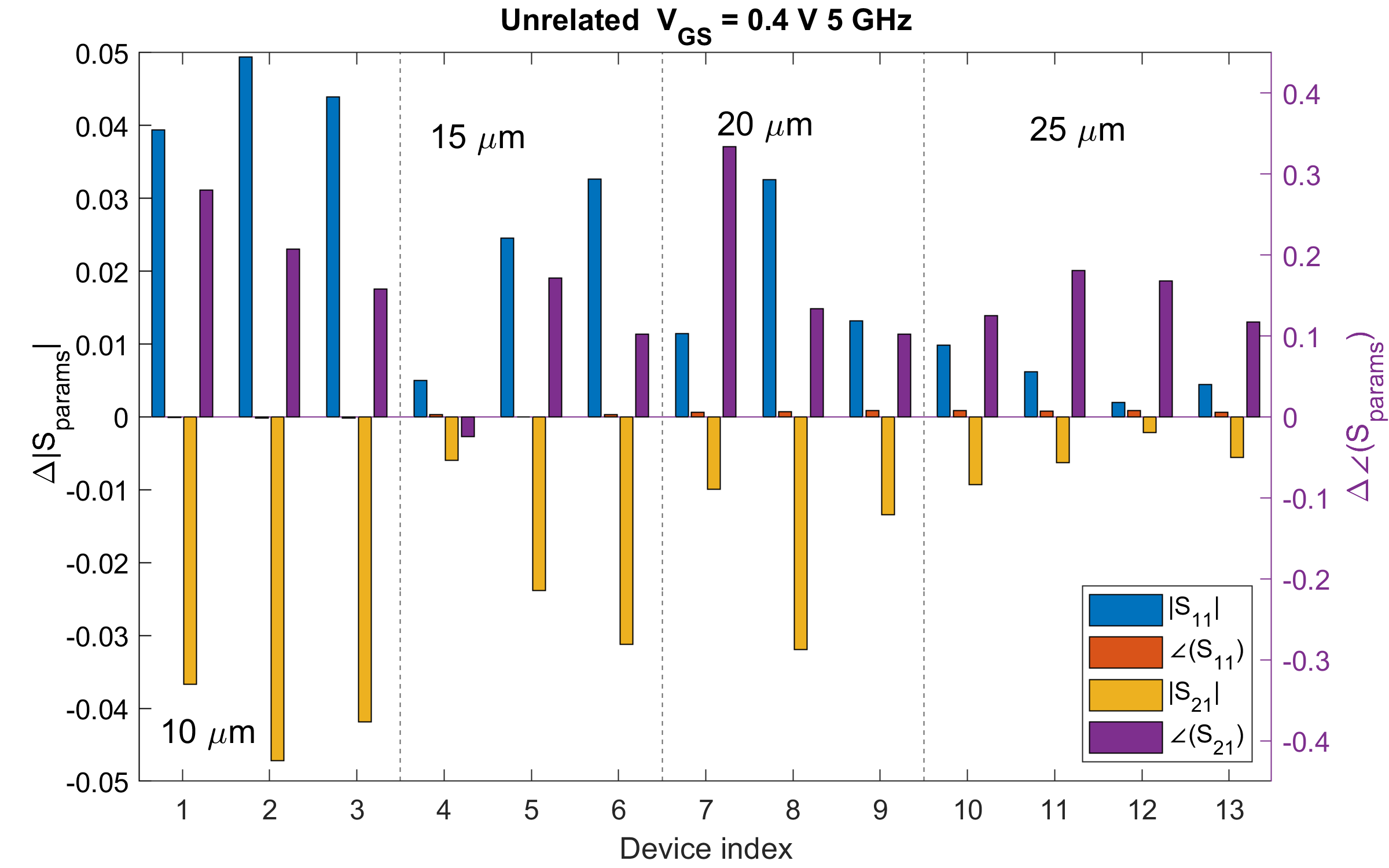}
\label{fig:AMP_UM_13devices}}
\caption{Bar graphs of the differences in the real, imaginary, amplitude, phase of $S_{21}$ and $S_{11}$ between the PBS baseline and pmDNA, smDNA, uDNA, respectively. Each group of the three/four bars is the result of an individual device. The frequency is 5 GHz. DNA concentration is 1 a\textsc{m}. The different colours represent different $S-parameter$ parts. For better visualisation, the y-axis on the left applies to the blue and yellow (Real and magnitude) components, while the y-axis on the right is for the orange and purple (Imaginary and phase) components.}
\label{fig:S-parameters13devices}
\end{figure*}

\newpage

\subsection{S-parameters at different DNA concentrations}
\vspace{0.9cm}

\begin{figure*}[htbp]
\centering
\subfloat[]{\includegraphics[width=0.32\linewidth]{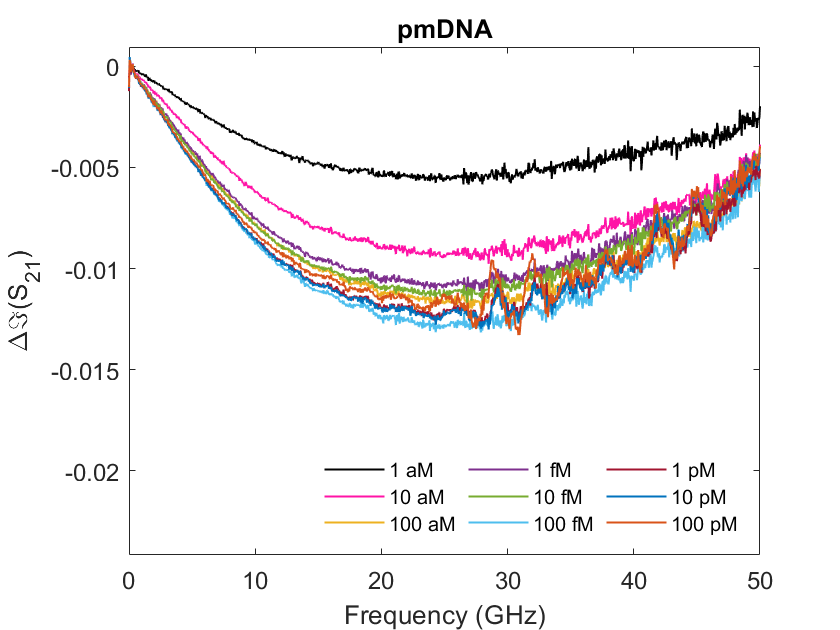}
\label{fig:pmDNA_9concentrations}}
\hfil
\subfloat[]{\includegraphics[width=0.32\linewidth]{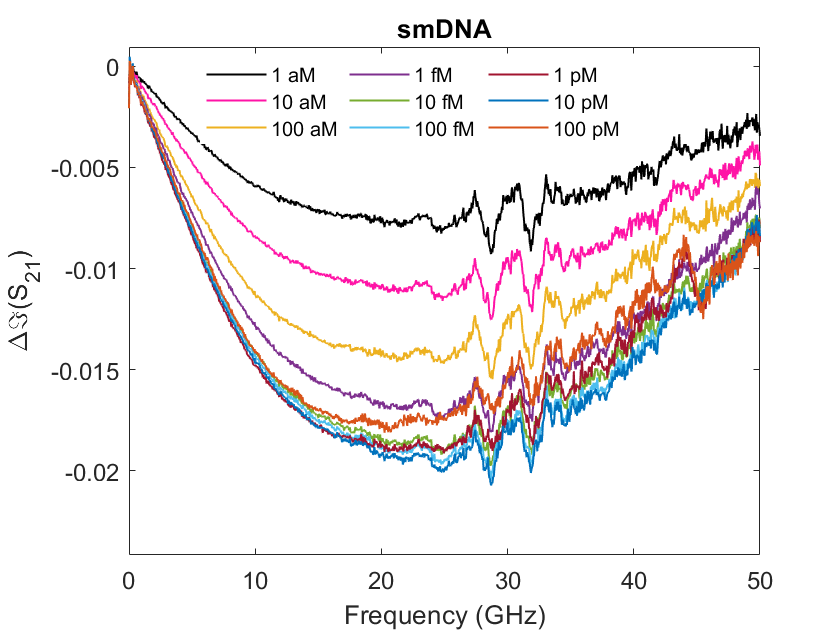}
\label{fig:smDNA_9concentrations}}
\subfloat[]{\includegraphics[width=0.32\linewidth]{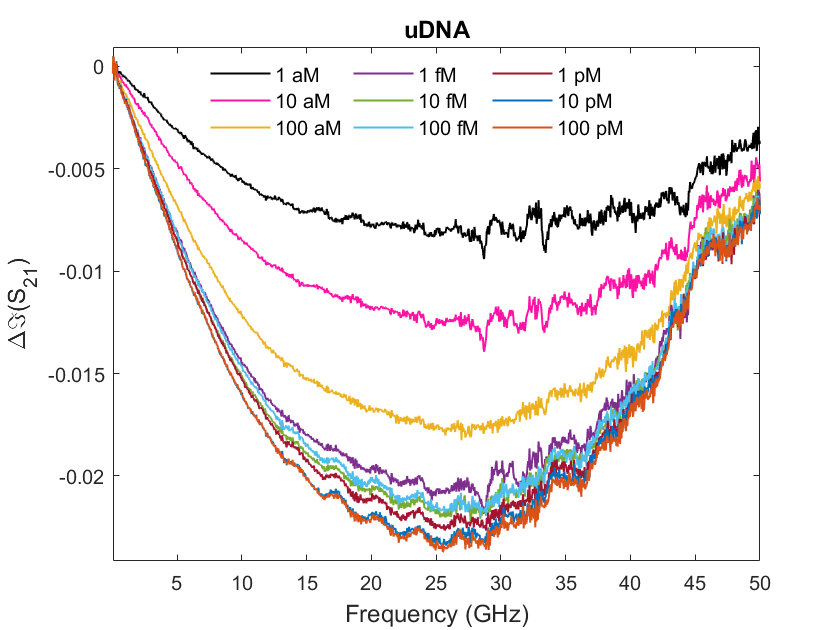}
\label{fig:uDNA_9concentrations}}
\hfil
\caption{A representative set of the difference in $\Im(S_{21})$ between the $\Im(S_{21})$ at nine different DNA concentrations and the $\Im(S_{21})$ at $0.01\times$ PBS for a representative device with a graphene channel length of $25 \mu m$ for (a) pmDNA (b) smDNA and (c) uDNA. Graphene channel length is 25 $\mu m$ at the gate voltage is 0 V. 
}
\label{fig:9concentrations}
\end{figure*}

\newpage

\section{Machine learning}
\begin{table}[!htbp]
\centering
  \caption{ML model classification accuracy of pmDNA at nine concentrations: 1 a\textsc{M}, 10 a\textsc{M}, 100 a\textsc{M}, 1 f\textsc{M}, 10 f\textsc{M}, 100 f\textsc{M}, 1 p\textsc{M}, 10 p\textsc{M}, and 100 p\textsc{M}. Whole dataset: 2583 (7 devices * 41 gate voltages * 9 concentration classes) $\times$ 4 (4 principal components), training set: 2070 $\times$ 4, testing set: 171 $\times$ 4.}
  \label{tbl:ClassificationResult_9concen}
  \begin{tabular}{lcc}
    \hline
    Algorithms & Accuracy (\%) & Accuracy with 
 30 dB Gaussian noise (\%) \\
    \hline
    LDA  & 100 & 100   \\
    SVM & 100 & 87.99 \\
    Tree & 99.75 & 98.04  \\
    KNN & 99.75 & 14.71 \\
    ANN & 100 & 100 \\
    \hline
  \end{tabular}
\end{table}

\begin{table}[!htbp]
\centering
  \caption{ML model classification accuracy of pmDNA, smDNA, and uDNA using the data at nine concentrations. Whole dataset: 6888 (7 devices * 41 gate voltages * 9 concentrations * 3 DNA classes) $\times$ 9 (9 principal components), training set: 5520 $\times$ 9, testing set: 1368 $\times$ 9.}
  \label{tbl:ClassificationResult_9concen_3classes}
  \begin{tabular}{lcc}
    \hline
    Algorithms & Accuracy (\%)\\
    \hline
    LDA  & 76.46   \\
    SVM & 85.67  \\
    Tree & 92.03  \\
    KNN & 100  \\
    ANN & 100  \\
    \hline
  \end{tabular}
\end{table}

\begin{figure}[h]
\centering
\subfloat[]{\includegraphics[width=0.48\linewidth]{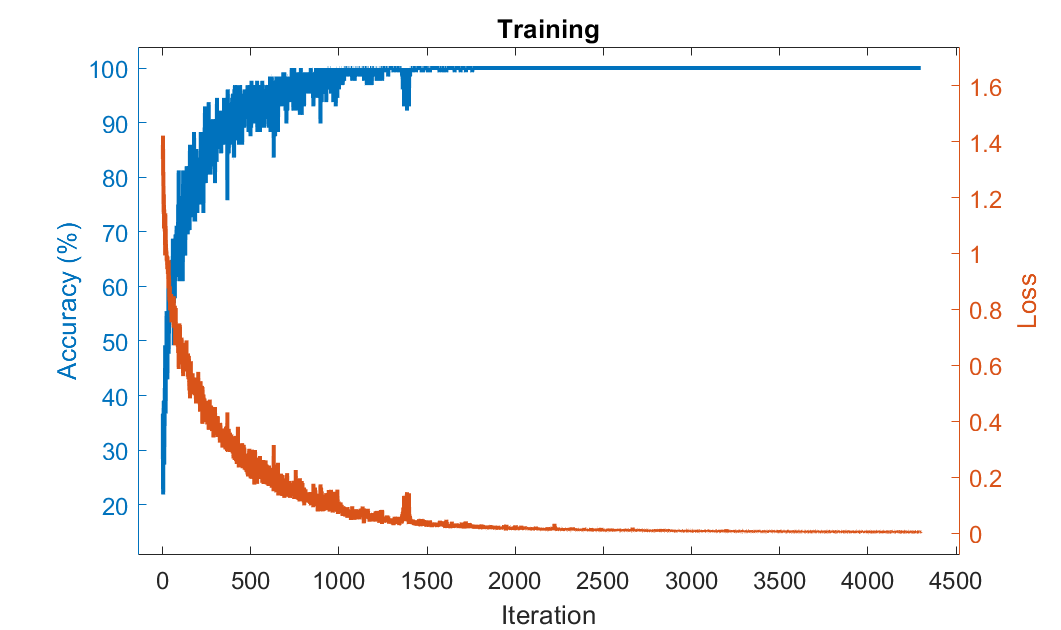}
\label{fig:training_9concen}}
\hfil
\subfloat[]{\includegraphics[width=0.48\linewidth]{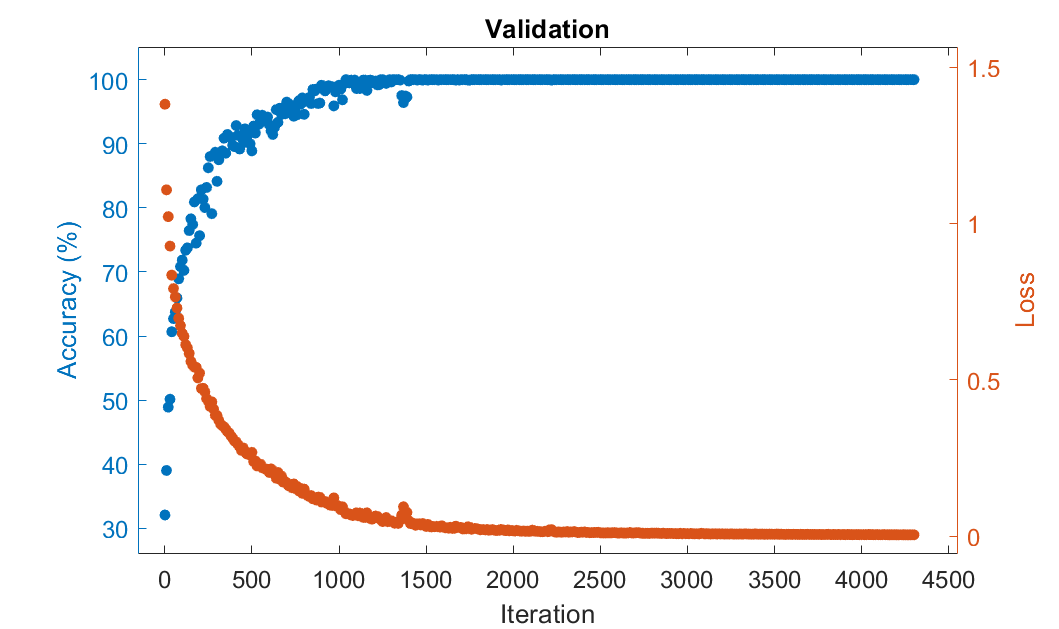}
\label{fig:validation_9concen}}
\caption{Accuracy and Loss curves of Training and Validation of ANN for the classification of pmDNA, smDNA, and uDNA using data at nine different concentrations. The validation accuracy is evaluated every ten iterations. Both validation and training accuracy increase with iterations and converge to 100\%. Therefore, the model is capable of classifying the three classes of DNA even at different concentrations.}
\label{fgr:ANN_9concen}
\end{figure}